 \documentstyle[12pt]{article}        
\begin{document}

\title{Quantum reference systems: a new 
framework for quantum mechanics}
\author{Gyula Bene\\
 Institute for Solid State Physics, E\"otv\"os University\\
M\'uzeum krt. 6-8, H-1088 Budapest, Hungary}

\date{\today}
\maketitle

\begin{abstract}
The new concept of {\em quantum reference systems} is introduced,
and a corresponding 
new foundation of nonrelativistic quantum mechanics is given in terms
of a set of postulates. The resulting theory gives an explanation for
 the measurement without assuming
 an {\em a priori} classical background.
  Schr\"odinger's cat paradox and the Einstein-Podolsky-Rosen 
  paradox gain resolution, too. 
  It is also shown that, despite of the 
 violation of Bell's inequality, quantum mechanics is a {\em local} 
 theory. 
\end{abstract}

\section{ Introduction: the problems}

Quantum mechanics proved to be extremely successful in describing
natural phenomena at the microscopic 
(molecular, atomic, subatomic) level. Moreover, there is no
evidence that its validity would
be limited when it is applied to macroscopic systems. Just
oppositely, in several cases it gives excellent 
results in solid state physics and the discrepancies in other
cases may be attributed to computational
difficulties rather than to some limitations of the theory
itself. Hence there is little doubt concerning the
universal validity of quantum mechanics, as long as nonrelativistic 
phenomena are concerned, and in the present paper
we also insist on this opinion. 

However, the basic laws of quantum mechanics, especially the 
superposition principle, seem to 
contradict sharply to the elementary observations concerning
macroscopic bodies. The observations 
(and even everyday experience) show that macroscopic bodies at 
a given instant of time have well 
determined coordinates and momenta (up to a minor inaccuracy
required by the uncertainty relations, which
is usually much smaller than the resolution of the available 
experimental techniques). Thus we may assume that
macroscopic systems should be quantum mechanically described by
such wave functions which are well 
localized both in coordinate and in momentum space. As an
example we may think of a narrow Gaussian 
wave packet
\begin{eqnarray}
\frac{1}{\sqrt[4]{2\pi \sigma^2}}\exp\left(-\frac{(x-\bar
x)^2}{4\sigma^2} +\frac{i}{\hbar} \bar p (x-\bar x)\right) \label{e1}
\end{eqnarray}
decribing the initial state of a point particle of large mass in one
dimension. Provided that $\sigma $ is much 
smaller than the characteristic
length scale on which the potential changes considerably and 
that $\frac{\hbar}{\sigma} $ is much smaller
than the expectation value of the momentum ($\bar p$), the wave 
packet is and remains to be localized
both in momentum and in coordinate space for a long time. 
Suppose now that $\Psi_1$ and $\Psi_2$ are such
localized states of the same macroscopic system, and they are 
macroscopically different, e.g. the distance
between the centers of mass of the wave functions is one meter. 
Quantum mechanics, according to the
superposition principle, would also allow the state
 $\alpha \Psi_1 + \beta \Psi_2$ to exist, where
$\alpha$ and $\beta$ are some nonzero complex numbers 
fulfilling $|\alpha|^2+|\beta|^2=1$. Such a state, however,
cannot be observed. 

Another puzzle can be that wavepackets in quantum mechanics
usually spread with time, thus after sufficiently
long time no localized states could be observed. 
This again contradicts the experience.

Thinking of Ehrenfest's theorem\cite{Ehr}, one suspects that the 
answer to
both questions can be that the initial
wave functions have been localized\footnote{Localization 
means here simply that the width of the wave function
at the given instant of time 
is small compared to the characteristic classical lengths in 
coordinate representation 
and is also small compared to the characteristic momentum in 
momentum representation. {\em It has nothing to do with Anderson 
localization.}}, and as 
the spread of the wave packets is extremely slow for a free particle 
of big mass (one gramm, say), the age of the universe has not
been enough for a considerable spreading. 
But this expectation is wrong! Let alone the question why the
initial conditions have been localized,
there are situations, when the spread of the wave packet is fast
even in case of macroscopic systems.

\begin{flushleft}
{\em i, Chaotic systems}
\end{flushleft}

This happens in classically chaotic systems\cite{chaos}, 
where wave packets spread 
exponentially with time in the semiclassical limit,
corresponding to the exponential separation of nearby 
 classical orbits. Therefore, in case of chaotic systems the time
needed for a spreading to macroscopically
observable sizes can be estimated by
\begin{eqnarray}
T=\frac{1}{\lambda}\ln\left(\frac{s_{final}}{s_{initial}}\right)\quad,
 \label{e2}
\end{eqnarray}
where $\lambda$ stands for the largest Lyapunov exponent
(a classical quantity!)\cite{Lyap}, while the widths 
of the wave packet
in the initial and in the final state are denoted by $s_{initial}$
and $s_{final}$, respectively. Obviously,
even for, say, $\frac{s_{final}}{s_{initial}}=10^{15}$ the time $T$ 
remains to be experimentally accessible. 
Chaotic systems, however, seem to be also well localized both in
coordinate and in momentum space 
(at any given instant of time). 

\begin{flushleft}
{\em ii, Quantum measurements}
\end{flushleft}

Another situation, where the explanation relying on
appropriate initial conditions fails,
is a typical quantum measurement when there are several possible
outcomes. Suppose, e.g., that the
 $z$-component of the spin ($\hat S_z$) of a spin-half particle $P$ 
 is measured by a suitable measuring device $M$, 
e.g. by a Stern-Gerlach apparatus. 
 If the state of the particle is an eigenstate of $\hat S_z$ 
($|\uparrow>$, say, 
corresponding to $S_z=+\frac{1}{2}$), then the measuring device
goes over into
 a state $|m_{\uparrow}>$ with unit probability. The state 
$|m_{\uparrow}>$, describing the situation 
when there is a spot on the upper side of the photograpic plate 
of the device is assumed 
to fit the experience, i.e., it is well localized both in
coordinate and in momentum representation.
The measurement process in this case may be symbolically written as
\begin{eqnarray}
|\uparrow>|m_0>\quad \rightarrow \quad |\uparrow>|m_{\uparrow}> 
\quad,  
 \label{e3}
\end{eqnarray}
where $|m_0>$ is the state of the measuring device before the
measurement (no spot on the photographic plate),
and $\rightarrow$ is a shorthand notation 
for the unitary time evolution during
the measurement, which is assumed to 
fulfill the time dependent Schr\"odinger equation. If the
initial state of the particle has been  
$|\downarrow>$ corresponding to $S_z=-\frac{1}{2}$, 
then the measurement process will be 
(analogously to Eq.(\ref{e3})) 
\begin{eqnarray}
|\downarrow>|m_0>\quad \rightarrow \quad |\downarrow>
|m_{\downarrow}> \quad.  
 \label{e4}
\end{eqnarray}
Suppose now that the initial state of the particle is not 
an eigenstate of 
$\hat S_z$, but a superposition of both, 
$\alpha |\uparrow>+\beta |\downarrow>$. According to the linearity
of the Schr\"odinger equation, the measurement process can be 
written now as
\begin{eqnarray}
(\alpha |\uparrow>+\beta |\downarrow>)|m_0>
\rightarrow \quad |\Psi>=\alpha |\uparrow>|m_{\uparrow}>
+\beta |\downarrow>|m_{\downarrow}>\quad.
 \label{e5}
\end{eqnarray}
One can see, that quantum mechanics predicts the final state of
the compound system $P+M$ to be 
a superposition of two, macroscopically 
distinguishable states, and this final state occurs with unit
probability. In contrast, in this situation one would 
actually observe that the state of the measuring device is
either $|m_{\uparrow}>$ or $|m_{\downarrow}>$,
and repeating the measurement under the same circumstances 
(i.e., with the same initial state), 
these possibilities would occur randomly, sometimes the first
one (with a probability $|\alpha|^2$), sometimes
the second one (with a probability $|\beta|^2$). By now it is 
well known and proven, that this randomness
cannot be a consequence of 
neglecting some local hidden parameters\cite{von Neumann}, \cite{hid} which
would allow for a deterministic description at a deeper
level\footnote{ Note that 
{\em nonlocal} hidden parameter theories can be constructed
\cite{nonloc}, however,
in the present paper we assume that the principle of locality 
never breaks down. }. 
 Thus, the above contradiction needs an
explanation within quantum mechanics. Note that the apparent
absence of macroscopic superpositions is expressed in 
an extreme form in Schr\"odinger's cat paradox\cite{cat}.

The problems described above (usually mentioned under the heading 
'the interpretation of quantum mechanics') 
has been unsolved for a long time, and during the many years 
a number of different attempts has been made for their solution. 
We apologize for not reviewing most of these. We do so
partly because of lack of space and partly
because the present approach is a completely new one.
Nevertheless, we shall briefly review
the Copenhagen interpretation, as this is the best known 
and most widely accepted
interpretation of quantum mechanics, and also, 
because the Einstein-Podolsky-Rosen (EPR) 
paradox relies on this interpretation. 

The relation between the present
interpretation and the so called decoherence theories will be discussed, too,
as the latter are quite popular today and determine the thinking of many
experts. The common element is the assumption that Schr\"odinger's 
equation is universally valid, implying that the transition from a pure
to a mixed state is a result of the interaction of 
the system with its environment. The difference is that decoherence theories 
concentrate mainly on the quantitative description of that transition
(i.e.,  on the effective solution of Schr\"odinger's equation written down
for a macroscopic system and its environment), while the present
interpretation concentrates on the meaning of quantum states, especially, 
on the meaning of the above mixed state and on its relation to the experience.

We shall also mention 
Everett's interpretation\cite{everett},
as it is based  on  similar assumptions as our approach 
(e.g., relativity of the
states,  
universality of Schr\"odinger's equation), and therefore 
there is a superficial similarity between the two approaches. 
This makes necessary 
to emphasize the rather fundamental differences, 
clearly distinguishing thus the present approach 
from Everett's interpretation. The distinction from two further 
approaches\cite{zeh},\cite{modal} will be discussed as well.

At this point the author feels it appropriate to sketch some
general features of his approach, in order to prepare the
reader for the coming hard trials. The difficulties in
understanding will be of conceptual rather than of mathematical
nature.\footnote{The author confesses
that he himself has had a very hard time to get used to the new
concepts, too.} The basic idea (cf. Section 3) itself will be
neither obvious nor natural, and the related postulates (cf.
Section 4) of the theory will look rather complicated and
arbitrary. Although some motivation and reasoning will be given
beforehand, the actual justification of the whole scheme comes
{\em a posteriori}. It is of worth emphasizing that the
correctness of a theory is independent of the emotions rised by
its strange-looking concepts. Instead, the relevant questions
are whether
\begin{enumerate}
\item the theory is free from internal contradictions,

\item its predictions are consistent with experience and  

\item with firmly established physical principles.
\end{enumerate}

As for question 1, the author shall try to convince the
reader that in case of the present theory the answer is affirmative.
Certainly, no rigorous proof of that will be given. 

As for question 2, it will be shown that the theory gives the
same theoretical predictions concerning the results of measurements as
traditional quantum mechanics, and in this sense the answer is 
affirmative. There are, however, other aspects as well:
if one applies traditional quantum mechanics to macroscopic systems,
one gets results which contradict the experience (Schr\"odinger's 
cat paradox). In the present theory no such contradiction
appears. 

Finally, traditional quantum mechanics precludes any
explanation of the classical properties, these are taken for granted.
 Our approach does not use the concept of a classical 
background,
therefore, if correct, it should lead under realistic conditions
to the observed macroscopic properties, like localization in
coordinate and in momentum space. The ultimate reason for such
a behavior is the interaction of the system with its environment
(which is another physical system). As the calculation scheme is 
well defined, the theory can be checked in numerical experiments.
 Whether the prescription
of the present theory indeed gives a correct result, is not known
at the moment and is not discussed in this paper.
 The question
of the emergence of classical properties has been considered by many
other authors, but the results and explanations strongly depend on the 
underlying interpretation. Papers based on the Schmidt decomposition\cite{sing} 
are relevant from the point of view of our approach as 
well\cite{zeh},\cite{modal},\cite{Schmidt},\cite{albrecht}.

Concerning question 3 one may tell that a distinctive feature
of the present interpretation is that according to it quantum mechanics
is consistent with the principle of locality.\footnote{This also 
implies that the present interpretation
cannot be considered as a 'rewording' of traditional quantum mechanics or
of any other previous interpretation.} This may be surprising to many,
as it is widely believed that the experimentally proven 
violation of Bell's inequality implies that either locality, or realism
or inductive inference is violated in Nature. Nevertheless, 
there is a fourth, evident-looking
assumption made at the derivation of Bell's inequality, namely, that if
two different states exist, they can also be compared, without destroying them.
The present interpretation shows that if two different states individually 
exist, it means that each of them may be checked by a suitable measurement 
without changing that state\footnote{But only {\em one} state may be actually 
checked
at the same time!}, but such a measurement will typically change
the other state. Therefore, these states, although individually exist,
are not comparable. Hence,
the violation of Bell's inequality implies the invalidity of the above 
fourth assumption rather than that of the principle of locality or realism
or inductive inference. One may claim that the answer to question 3 is 
positive, too.

Some other important features of the present theory are
the following:
\begin{enumerate}
\item the Schr\"odinger equation is assumed to be universally valid
 (if relativistic effects are negligible),
 
\item  measurements are explained in terms
of the usual interactions,

\item there is no 'reduction' or 'collapse' of the wave function,

\item the new formulation accomodates the indeterministic
nature of quantum mechanics,

\item all the conclusions can and should be drawn by using
exclusively the rules of the present theory. Handwaving, intuitive
arguments or arguments borrowed from other interpretations
are not appropriate.
\end{enumerate} 

The paper is organized as follows. 
In Section 2 we briefly review and criticize the 
Copenhagen interpretation and in particular the concept of
the collapse of the wave function, aimed at solving the problem
of the measurement. The motivation for and the essence of 
the basic new physical assumption of the 
paper is given in Section 3. 
We introduce here the concept of the
{\em quantum reference systems} on which quantum states depend. 
The differences from the above mentioned other interpretations
are discussed here, too.
The new physical assumption gives a 'new freedom' 
in the theory, and makes necessary a development of a 
new formulation of quantum 
mechanics giving account of the relation among states with respect to
different reference systems. This is done in Section 4 such a
way that all the well established results of quantum mechanics
remain unchanged. In Section 5 Schr\"odinger's cat paradox is 
discussed and it is demonstrated that within the present approach 
the paradox disappears. In Section 6 we discuss
the Einstein-Podolsky-Rosen paradox and 
show that it gains resolution
within the new framework. In Section 7 
the violation of Bell's inequality is analyzed. It is shown that
according to the present approach 
correlations between measurements done on separate particles 
can be attributed exclusively to a previous interaction 
('common past')
 of the particles, thus quantum mechanics is a {\em local} theory.
In the concluding Section 8 we summarize the results and 
 discuss some general features 
of our approach like unitary covariance of the formulation and
the relation of quantum reference systems to the usual ones.
A derivation of the so called Schmidt canonical form is given 
in Appendix A. 
Appendix B discusses the relation of the present approach to 
the Copenhagen
interpretation, showing that the latter corresponds to 
a special choice of the quantum reference system.
\vskip0.5cm
\section{ Criticism of the traditional (Copenhagen) solution of the
problem} 

Traditionally, one assumes that 'due to the macroscopic nature' 
of the system $P+M$ the wave function
'collapses' either (with probability $|\alpha|^2$) 
to $|\uparrow>|m_{\uparrow}>$ or 
(with probability $|\beta|^2$) to
$|\downarrow>|m_{\downarrow}>$\cite{Landau}. 
As is well known, using this assumption
one may construct a paradox, discovered by Einstein, Podolsky
and Rosen (EPR in a shorthand notation) \cite{EPR}. 
These authors considered
a situation when two separate particles are described by a
correlated wave function (i.e., with a
sum of product states, instead of a single product of the states
of the particles), due to
some interaction having taken place in the past. If a
measurement is made on the first particle,
and the collapse of the wave function of the whole system takes 
place, then one concludes
that the state of the second particle becomes a definite wave
function, and it depends
on the kind and result of the measurement, although the second
particle does not interact
with the measuring device, nor with the first particle during 
and after the
measurement. 
The original conclusion drawn by EPR
was that quantum mechanics did not give a complete description
of the physical processes. This 
conclusion, in view of the developments since, especially of 
the failure of local hidden variable theories\cite{hid},
cannot be accepted. The paradox, however, should be 
somehow resolved. If one wants to maintain 
the concept of the collapse of the wave function there are 
essentially two possibilities. 
\vskip0.5cm
{\em a}. It is well known that the EPR paradox does not imply that 
quantum mechanics contradicts the experiments.
It points out the ambiguity of the concept of the wave function.
If one does not assume that the wave
function corresponds to some state which actually exists (i.e., which 
would be an 'element of the reality', using EPR's 
terminology), then no contradiction appears. This leads 
to the only consistent version of the Copenhagen
 interpretation: the wave functions have no objective meaning, 
they are simply calculational tools\cite{Wigner1}.
 The aim of the theory is the prediction of the correlations 
between successive measurements, and the 
 only 'elements of reality' are the results of the measurements.
\vskip0.5cm 
{\em b}. One assumes that the wave functions correspond to 
some objective states of the systems. Then the EPR paradox implies
the existence of an 'instantaneous' interaction, which is
faster than the light, violating also causality, although
it does not appear directly in experiments. 

Experiments actually prove a
correlation between separated, noninteracting particles.
One would think that this is a consequence of the previous interaction
('common past') between them. If, however, one pursues this
line, some inequalities are found for the joint
probabilities of the measurements done on the two
particles\cite{hid}. As is known, Bell's inequality is
not always fulfilled by quantum mechanics, moreover,
experiments support the quantum mechanical prediction\cite{exp}.
As it has been stressed by d'Espagnat\cite{Esp}, the conclusion would
be, as long as one is to assume the existence of some
objective states (not necessarily described by wave
functions)\footnote{I.e., using the terminology of Ref.\cite{Esp},
when we have to do with a realistically interpretable theory.}, 
 that the world is 'nonlocal', as separated
particles can 'feel' each other. 
\vskip0.5cm
The possibility {\em a}, although is now widely accepted, does
not seem to be satisfactory (among others, in quantum
cosmology). It is hard to believe that quantum mechanics 
is not a model of the physical world and that the only reality is 
the result of a measurement. One is inclined to reject
the collapse of the wave function, rather than confine
physics to the laboratories. 

As for the possibility {\em b}, it does not seem to be
acceptable to reject such well proven physical principles,
like causality and locality. Nevertheless, it is less obvious
how this problem is related to the collapse of
the wave function. One may think that
rejecting the existence of the collapse the original EPR
argument will not work, but the observed correlations
and the violation of Bell's inequality remain there
and may still imply 'nonlocality'. We shall see, however,
that the proposal described below solves this difficulty,
too, as the derivation of Bell's inequality contains 
a hidden assumption, not allowed according to the proposed theory.
Thus locality will be restored and the observed
correlations will be completely attributed to the
interaction having taken place in the past. 
\vskip0.5cm
\section{The basic idea of the proposed solution}

In view of the difficulties caused by the collapse of the
wave function we reject its existence, but keep all the
other parts of quantum mechanics (including Schr\"odinger's
equation, the expression for the probabilities of the
possible results of a measurement etc.). Of course, we are
then back at the contradiction presented in Section 1: the
quantum mechanical prediction is one definite state, a
superposition of two localized states, but experience
corresponds to one or the other localized state. As a first
remark, let us observe that experience refers to the state
of the measuring device $M$ rather than that of the whole 
system $P+M$\footnote{Quite rigorously, experience refers
to the content of the mind of an observer \cite{Wigner2}, and
what follows will not contradict this view, however, such a
rigour which would rise too much philosophical questions is
not needed here.}, thus we have to compare prediction and
experience concerning the state of the measuring device. As the
measuring device is a subsystem having no own wave function
one has to use a {\em reduced density matrix}\cite{dens} for
its description. It is defined by
\begin{eqnarray}
\hat \rho_M=Tr_P\left(|\Psi><\Psi|\right)\quad,
 \label{e6}
\end{eqnarray}
where $Tr_P$ stands for the trace operation in the Hilbert
space of the particle $P$. (As the Hilbert space built 
up of all the possible states of a physical system is uniquely 
related to that system, here and hereafter we use the term 
'the system's Hilbert space'.)
As is known, this object contains all the
information necessary to predict the probabilities of the
possible results of any measurement done on the subsystem $M$.
Inserting the expression for $|\Psi>$ (cf. Eq.(\ref{e5}))
 into Eq.(\ref{e6}) we get 
\begin{eqnarray}
\hat \rho_M=|m_{\uparrow}>|\alpha|^2<m_{\uparrow}|
+ |m_{\downarrow}>|\beta|^2<m_{\downarrow}|\quad.
 \label{e7}
\end{eqnarray}
Comparing it with the experience one can see that the
actually observed states of the measuring device are just
the eigenstates of the reduced density matrix $\hat \rho_M$, and
their probabilities to occur are the corresponding
eigenvalues. Nevertheless, despite of this remarkable
finding, we still have the contradiction: the quantum mechanical
prediction is the whole reduced density matrix, while only
one of its eigenstates is observed. According to our
assumptions, we accept that the quantum prediction is
correct without any modification (collapse, e.g.), and, of
course, we also accept the experience. 
 In order to resolve the apparent contradiction, let us discuss
  a bit more detail how we have got the quantum prediction and what it means.
  When calculating $\hat \rho_M$, we have used the wave function $|\Psi >$
  (cf. Eq.(\ref{e5})) of the whole system $P+M$. Traditionally a wave
  function is conceived as the result of a suitable measurement (the
  preparation), thus one can tell that when calculating $\hat \rho_M$
  we have actually used the information gained from a measurement
  done on the whole system $P+M$. In contrast, in case of the experience,
  a 'measurement', i.e., an observation is done directly on the system $M$.
  Therefore, the reason for the difference between prediction and
  experience is that the information used when describing
  the system $M$ stems from measurements done on different
  systems.
 Let us call the system which has been measured (it is $P+M$
in the first case and $M$ in the second case) the {\em quantum
reference system}. Using this terminology, we may tell that
we make a measurement on the quantum reference system $R$, thus we prepare
its state $|\psi_R>$ and using this information we calculate
the state $\hat \rho_S(R)=Tr_{R\setminus S} |\psi_R><\psi_R|$ 
of a subsystem $S$. We shall call
$\hat \rho_S(R)$  the state of $S$ with respect to $R$. 
Obviously $\hat \rho_R(R)=|\psi_R><\psi_R|$, thus $|\psi_R>$
may be identified with the state of the system $R$ 
with respect to itself. For brevity we shall call this the internal state of $R$.
 
Let us emphasize that up to
this point, despite of the new terminology, nothing has been
added to usual quantum mechanics. 

Let us return now to the question why 
the state of a system $S$ depends on the 
  choice of the quantum reference system $R$ ($S\subseteq R$). It is certainly very surprising
from the classical point of view that the state of a system
is found to be different if we measure the system directly or if
we measure it together with an environment. Nevertheless, in the spirit
of the Copenhagen interpretation one
may tell that this is just because in quantum mechanics
measurements usually change the states, different measurements lead to
different changes, thus the above mentioned
difference may be attributed \newpage
completely to the different measurements. 
This argument is, however, not compelling.
 At this decisive point we leave the traditional framework of
 quantum mechanics and make the fundamental assumption that
 the dependence of the states on 
quantum reference systems is an inherent property of quantum mechanics 
and not the result of the disturbance due to measurements. At the same
time it is also assumed 
that the states exist even in the absence of any measurement\footnote{
Of course, these states are unknown for us in the absence of measurements 
and usually will be changed if this or that 
measurement is performed.}. 
Our assumption means that one and the same system $A$ 
is characterized by a multitude
of states $\hat \rho_A(R_1),\;\hat \rho_A(R_2),\;...$, each referring to
a different quantum reference system $R_i$. As for their physical
meaning, one can conceive the state $\hat \rho_A(R)$ as the 
description of the system $A$ 
using the information gained from a suitable measurement
performed on $R$. The term 'suitable' means that the measurement 
does not change the state $\hat \rho_R(R)$  (the state of the
quantum reference system with respect to itself), otherwise the state
$\hat \rho_A(R)$ changes due to the measurement. 
Let us draw the attention of the reader
to a subtle point: although one cannot know {\em a priori}
what such a suitable measurement is, it does not lead to any
contradiction if we assume the existence of such a measurement. 
 The possibility of nondisturbing measurements is
the expression of realism: the states exist whether we measure it or not,
and in principle they can be learned.

Note that the above assumption is a radical departure from 
the conventional point of view of  quantum mechanics. There
anything that exists is associated with 
some actual measurement, thus states themselves are thought to be
created by measurements. Obviously, such an approach excludes from the 
beginning the possibility to describe the measurements in terms of
quantum mechanics, as measurements are treated as primary entities
and states as the secondary ones. (Even if one considers the measuring device
as part of the quantum system, a further measuring device and measurement is
needed to describe the states according to the usual 
Copenhagen interpretation.\cite{von Neumann})
In contrast, in our approach the primary concepts are the states,
and measurements are derived. As noted above, this is the obvious
prerequisite if one wants to describe quantum measurements 
in terms of quantum mechanics. 
We shall see that in terms of the states depending 
  on quantum reference systems one can actually interpret the
  measurements independently of classical mechanics and can establish 
  quantum mechanics consistently, freely from paradoxes.
  
  Let us summarize the main idea expounded above:
  \vskip0.5cm
{\em The dependence of the state of a physical system on quantum 
reference systems is an inherent property of quantum mechanics.
Consequently, a state with respect to the chosen quantum
reference system exists even in the absence of measurements, and 
in principle there
exist a suitable measurement which, if
 performed on the quantum reference system in order to determine this state, does not disturbe it.} 
\vskip0.5cm
This conclusion is the essential new physical assumption
upon which the whole theory presented in this paper is
based. Let us discuss it in a bit more detail. 

The meaning of the quantum reference systems is rather analogous
to the classical coordinate systems. Choosing a
classical coordinate system means that we imagine
what we would experience if we were there. Similarly,
choosing a quantum reference system $R$ means that we
imagine what we would experience if we did a measurement on
$R$ that does not disturbe $\hat \rho_R(R)=|\psi_R><\psi_R|$. 
In order to see that such a measurement exists, consider an
operator $\hat A$ (which acts on the Hilbert space of $R$) whose
eigenstates include $|\psi_R>$. The measurement of $\hat A$
will not disturbe $|\psi_R>$. 

Nevertheless, there exist important features of the 
quantum reference systems that are quite unusual 
from the classical point of view. It is certainly not surprising,
that a given system can be characterized by a multitude of states,
each referring to a different quantum reference system.
One may safely say that all these states exist, just like 
in case of the classical coordinate systems all the states 
with respect to different coordinate systems exist.
Indeed, in the spirit of Einstein, Podolsky and Rosen we may accept
that a state $\hat \rho_S(R)$ is an element of the reality if
there exists a suitable nondisturbing measurement to it. 
As we have seen, such a measurement always exists. The nondisturbing
measurement may of course be repeated and one gets the same result, as before.
Nevertheless, the states defined
with respect to different quantum reference systems are not
necessarily comparable. What does it mean? 
We explain it on an explicit example. Let us consider three distinguishable
spin-half particles, $A$, $B$ and $C$. Be $R_1=A+B$ and $R_2=B+C$.
Suppose that the system $A+B+C$ is isolated. Denoting
the eigenstates of the $\hat S_z$, $\hat S_x$ spin-component by
$|\uparrow>$, $|\downarrow>$ and $|+>$, $|->$, respectively, be the state of
$A+B+C$ (considering only the spin degrees of freedom)
\begin{eqnarray}
|\phi>=\alpha\left(\beta |A,\uparrow>|B,\downarrow>
+\gamma^* |A,\downarrow>|B,\uparrow>\right)|C,+>\nonumber\\
+\delta\left(\gamma |A,\uparrow>|B,\downarrow>
-\beta^* |A,\downarrow>|B,\uparrow>\right)|C,->\\
=|A,\uparrow>|B,\downarrow>\left(\alpha \beta |C,+> +
\delta \gamma |C,->\right) \nonumber\\
+|A,\downarrow>|B,\uparrow>\left(\alpha \gamma^* |C,+> -
\delta \beta^* |C,->\right)\nonumber
\end{eqnarray}
with
$|\alpha|^2+|\delta|^2=1$, $|\beta|^2+|\gamma|^2=1$. 
In order to calculate the states $\hat \rho_{R_1}(R_1)$ 
and $\hat \rho_{R_2}(R_2)$ one has to calculate the density matrices
$\hat \rho_{A+B}(A+B+C)$ and $\hat \rho_{B+C}(A+B+C)$,
respectively, and has to find the eigenvectors (cf. Section 4, 
{\em Postulates 4, 6}
and {\em Proposition 1}). The result is that $\hat \rho_{R_1}(R_1)$ is
a projector corresponding either to the state vector
\begin{eqnarray}
|\psi_+>=\beta |A,\uparrow>|B,\downarrow>
+\gamma^* |A,\downarrow>|B,\uparrow>
\end{eqnarray}
or to
\begin{eqnarray}
|\psi_->=\gamma |A,\uparrow>|B,\downarrow>
-\beta^* |A,\downarrow>|B,\uparrow>\;,
\end{eqnarray} 
while $\hat \rho_{R_2}(R_2)$ is
the projector corresponding either to 
\begin{eqnarray}
\left(|\alpha|^2|\beta|^2+|\delta|^2|\gamma|^2\right)^{-\frac{1}{2}}
|B,\downarrow>\left(\alpha \beta |C,+> +
\delta \gamma |C,->\right)\label{bels1}
\end{eqnarray}
or to
\begin{eqnarray}
\left(|\alpha|^2|\gamma|^2+|\delta|^2|\beta|^2\right)^{-\frac{1}{2}}
|B,\uparrow>\left(\alpha \gamma^* |C,+> -
\delta \beta^* |C,->\right)\;.\label{bels2}
\end{eqnarray}
Suppose we make a nondisturbing measurement on $R_1$. The dynamics
of this measurement may be written symbolically as
\begin{eqnarray}
|\psi_\pm>|m_0>\rightarrow |\psi_\pm>|m_\pm>\;.
\end{eqnarray}
This implies that after the measurement 
the state of the whole system $A+B+C+M$ is given by
\begin{eqnarray}
\alpha\left(\beta |A,\uparrow>|B,\downarrow>
+\gamma^* |A,\downarrow>|B,\uparrow>\right)|m_+>|C,+>
\mbox{\hspace{0.5cm}}\nonumber\\
+\delta\left(\gamma |A,\uparrow>|B,\downarrow>
-\beta^* |A,\downarrow>|B,\uparrow>\right)|m_->|C,->\;.
\end{eqnarray}
Repeating the calculation for $\hat \rho_{R_2}(R_2)$\footnote{Note 
that the state $\hat \rho_{R_1}(R_1)$ does not change, 
as expected.} one gets now the
result that it is a projector corresponding to one of the four
state vectors
\begin{eqnarray}
|B,\downarrow>|C,+>\\
|B,\downarrow>|C,->\\
|B,\uparrow>|C,+>\\
|B,\uparrow>|C,->\;.
\end{eqnarray}
Comparing these states with Eqs.(\ref{bels1}), (\ref{bels2}) one can see
that the measurement which did not disturbe $\hat \rho_{R_1}(R_1)$,
changed $\hat \rho_{R_2}(R_2)$. Therefore, we cannot learn
both $\hat \rho_{R_1}(R_1)$ and $\hat \rho_{R_2}(R_2)$ at the same time
without changing at least one of them, so we cannot compare these states. 
We can
learn only one of them, certainly, we may decide, which one. We shall express
this surprising property by saying that although both 
$\hat \rho_{R_1}(R_1)$ and $\hat \rho_{R_2}(R_2)$ exist, they are not 
comparable. It is important to understand, that this property does not
influence the realism of the theory. Mathematically we may conceive
reality as a set containing states and sets of states.
If a suitable nondisturbing measurement existed, so that 
both $\hat \rho_{R_1}(R_1)$ and $\hat \rho_{R_2}(R_2)$ could be
learned at the same time without changing them, it would mean
that not only $\hat \rho_{R_1}(R_1)$ and $\hat \rho_{R_2}(R_2)$ as individual
states, but also the set containing both states would be an element of reality.
Certainly if two states are elements of the reality, it does not imply
automatically
that the set of those two states is also an {\em element} of the reality.
It implies only that the set of the two states is a {\em subset} of reality.
Indeed, in the above example both $\hat \rho_{R_1}(R_1)$ 
and $\hat \rho_{R_2}(R_2)$ are
elements of reality as individual states (as there exists for each a suitable
nondisturbing measurement), but the set of these states is not
an element of reality, as no corresponding nondisturbing
measurement exists. 

It is interesting to note that there exists a classical analogue of the
above noncomparability, i.e., descriptions with respect to different
coordinate systems may be noncomparable.
 It is known that general
relativity allows coordinate systems even inside of a black
hole. Consider now two different black holes and
introduce a coordinate system inside each of them.
Probably no one doubts the reality of the descriptions
with respect to these coordinate systems. We may indeed
check what we may experience with respect to {\em one} of these
systems. We may freely choose one of the black holes andmay\hfill\break
fall into it. Then we see what is inside. However, if we do so
we cannot come back and thus automatically prevent ourselves
from learning the other blackhole interior. Therefore, 
analogously to the quantum case, descriptions with respect 
both to the one and to the other black hole interior exist, but they
cannot be observed simultaneously. Of course, we
do not want to say that there is any deeper connection between
the underlying physics of the quantum case and that of the above 
classical example. Just oppositely, the mechanisms leading to noncomparability 
are completely different in the classical and the quantum case.
In the quantum case a nondisturbing measurement performed on one system
changes the state of the other system. In contrast, in the classical case
performing an observation in one coordinate system prevents from performing
any observation in the other system. There is no disturbance on the
other coordinate system at all.

 An important feature of the usual coordinate 
systems is that
properties of a given physical system observed from 
different coordinate
frames are unambigously related, namely, quantum states 
with respect to
different coordinate systems are connected via suitable unitary
transformations. The concept of quantum reference systems
is different: usually  the relation of states 
with respect to different quantum reference systems is not one-to-one: 
in the above example, if the state of $M$ with respect to $P+M$ is
$\hat \rho_M$, then the state of $M$ with respect to $M$ can be 
either $|m_{\uparrow}>$,
or $|m_{\downarrow}>$. It is of worth mentioning that this
feature expresses the indeterminism of quantum mechanics in our approach.
Nevertheless, the concept of the
usual coordinate frames and that of the quantum reference systems
are connected: as we shall see, the former can be considered
a special case of the latter. From the mathematical point of view,
 however,
they are rather different: a classical coordinate system is a 
coordinate
system in the mathematical sense, while a quantum reference system is 
a Hilbert space.

We discuss now the (rather fundamental) differences 
between our approach and Everett's interpretation\cite{everett}. Everett 
writes about 'relative states' of a subsystem, and he defines them with
respect to the {\em states} of the complementary subsystem. In
contrast, in
our approach the state of a subsystem is defined with respect
to a {\em system} which contains the subsystem to be described.
This difference leads to a completely different physical picture
in the two cases: Everett attributes a physical meaning to the
product components of the entangled state of an isolated compound
system, all being somehow relevant (parallel worlds), while in
the present interpretation these components do not have
separately any physical meaning, only together, and this whole
entangled state describes the compound system with respect to
itself (i.e., the quantum reference system is the compound
system). The state of a subsystem with respect to this quantum
reference system is the corresponding reduced density matrix and
not some factor of the  component product states.
The problem why we see only one result at each quantum
measurement when the whole wave function of the universe is
still present is solved in the present interpretation by the
realization that in case of a quantum measurement the measuring device (or the
observer), i.e., a subsystem is the relevant 
quantum reference system. Therefore, the relevant 
quantum reference system is different than before, and there
is no logical necessity that a one-to-one relationship
between the descriptions with respect to these different
quantum reference systems exists. This remark also shows
that the present interpretation does not involve the parallel
worlds. 

Ideas somewhat similar to ours has been put forward in
Ref.\cite{zeh}, where the author writes about Schmidt states 
as experienced 'subjectively' by the observer. The concept of
quantum reference systems is,
however, not introduced there, instead, the idea mentioned above
is
associated with Everett's interpretation, which
differs markedly from our approach.

Another similar interpretation is the so called modal
interpretation\cite{modal}, which states that one of the eigenstates of the
reduced density matrix corresponds to an actually existing
property of the subsystem. ('Actually existing' here means that
in an absolute sense, independently of quantum reference
systems.) In our approach these states are the states of the
subsystem with respect to itself, i.e., they are not relevant
with respect to other quantum reference systems. In Section 7
we argue that the modal interpretation implies Bell's
inequality, while our approach does not. 

Finally, it is reasonable to discuss the relation of the present 
approach to the so called decoherence theories\cite{Zurek}, \cite{zeh}. 
These latter are quite popular today and determine the thinking of
many experts. The main idea is that the transition of a macroscopic system 
$M$ from a pure to a mixed state is due to the interaction
between the system $M$ and its environment $E$. The effect is
actually a consequence of the time-dependent Schr\"odinger equation 
written down for the closed $M+E$ compound system, i.e.,
decoherence theories assume the universality of the Schr\"odinger equation, 
just like the present approach. Explicitly, if initially $M$ is in a pure state
$|\varphi>$, and correspondingly the system $M+E$ is in the 
product state $|\varphi>|\xi>$, the time dependent Schr\"odinger
equation written down for $M+E$ leads to the unitary time evolution
\begin{eqnarray}
|\phi(t)>=\exp\left(-i\frac{\hat H t}{\hbar}\right)\left(|\varphi>|\xi>\right)
\end{eqnarray}
where the Hamiltonian $\hat H$ includes the interaction term between
$M$ and $E$. Due to this term $|\phi(t)>$ ceases to be of product form, 
instead, it becomes a sum of product states,
\begin{eqnarray}
|\phi(t)>=\sum_j c_j(t) |\varphi_j(t)>|\xi_j(t)>\,,\label{dc}
\end{eqnarray}
where $|\varphi_j(t)>$-s and $|\xi_j(t)>$-s are orthonormed sets of states
(at any given instant of time $t$). Note that the form (\ref{dc}),
called Schmidt decomposition (cf. Appendix A) does not restrict
the generality of the discussion. Calculating the reduced density matrix
from $|\phi(t)>$ for the system $M$ one obtains
\begin{eqnarray}
\hat \rho_M(t)=\sum_j |\varphi_j(t)>|c_j(t)|^2<\varphi_j(t)|\;.
\end{eqnarray}
The transition from a pure to a mixed state is actually a generic
property of interacting quantum systems (even if they are not
macroscopic). We also used it already in Eqs.(3)-(7). There the
macroscopic system was the measuring device and the role of the
environment was played by the spin-half particle. Therefore,
up to this point the present approach agrees with that of 
decoherence theories. The difference between the two approaches
lies in the different questions these theories try to answer. 
The main achievement of decoherence theories is the quantitative
description of the transition from a pure to a mixed state, i.e., 
the actual calculation of $\hat \rho_M(t)$ for various, more or less
realistic models. This is that part of decoherence theories where
one may speak about a consensus in the literature. Concerning the
physical meaning of $\hat \rho_M(t)$, i.e., its relation to the experience,
no such consensus has been achieved. The present approach puts emphasis
exactly on that problem. It will be shown that the new idea presented in this section
leads to a consistent set of postulates, which offer in principle an answer 
to all potential problems. The power of this approach will be demonstrated
in Sections 5-7, where Schr\"odinger's cat paradox and the EPR paradox
will be resolved and the violation of Bell's inequality will be explained
without giving up the principle of locality.
\vskip0.5cm
\section{Rules of the new framework}

In this section the new version 
of the rules of quantum mechanics is formulated, including 
the connections between states of a system with respect to
different reference systems. We shall see that the 
principle that quantum states depend 
on the quantum reference systems
can be incorporated into quantum mechanics so as to recover 
its usual results exactly.

Let us denote the physical system to be described by $A$ and the 
reference system by $R$. The state of $A$ with respect to $R$
will be denoted by $\hat \rho_A(R)$.
\vskip0.5cm
  {\em Postulate 1. The system A to be described is a subsystem of the
reference system R.} 
\vskip0.5cm 
Note that the reference system may coincide with the system to
be described ($A=R$). In such a case we speak about an {\em 
internal state}.
\vskip0.5cm
{\em Definition 1. $\hat \rho_A(A)$ is called the internal state 
of $A$.}
\vskip0.5cm
{\em Postulate 2. The state $\hat \rho_A(R)$ is a positive definite, 
Hermitian
operator with unit trace, acting on the Hilbert space of $A$.}
\vskip0.5cm
\vskip0.5cm
Note that the unit trace means that we confine our considerations
to normalizable states, which is not a serious restriction in 
nonrelativistic 
quantum mechanics. 
\vskip0.5cm
{\em Postulate 3. The internal states $\hat \rho_A(A)$ are always 
projectors, i.e., $\hat \rho_A(A)=|\psi_A><\psi_A|$.}
\vskip0.5cm
In what follows these projectors will be identified with the 
corresponding wave functions $|\psi_A>$ (as they are uniquely
related, apart from a phase factor). According to {\em Postulate 2} 
the internal states are normalized to unity.
\vskip0.5cm
{\em Postulate 4. The state of a system $A$ with respect to the
reference system $R$ (denoted by $\hat \rho_A(R)$) is the reduced
density matrix of $A$ calculated from the internal state of $R$,
i.e.
\begin{eqnarray}
\hat \rho_A(R)=Tr_{R\setminus A} \left(\hat \rho_R(R)
\right)\quad,
 \label{e8}
\end{eqnarray}
where $R\setminus A$ stands for the subsystem of $R$ 
complementer to $A$.}
\vskip0.5cm
One can see that {\em Postulate 4} is consistent with 
{\em Postulate 2} and {\em Postulate 3}.
\vskip0.5cm 
We introduce now the notion of {\em isolated } and  
{\em closed systems}.
\vskip0.5cm
{\em Definition 2. An isolated system is 
such a system that has not been interacting 
with the outside world. A closed system
is such a system that is not interacting with any other
system at the given instant of time 
(but might have interacted in the past).}
\vskip0.5cm 
An isolated system can be
described by a wave function, and, moreover, this wave function
always occurs as a factor in the internal state of any
broader system. Thus we set
\vskip0.5cm
{\em Postulate 5. If $I$ is an isolated system then its state is 
independent of the reference system $R$:}
\begin{eqnarray}
\hat \rho_I(R)=\hat \rho_I(I)\quad.
 \label{e9}
\end{eqnarray}
\vskip0.5cm
In other terms, {\em Postulate 5} means that the state of an isolated
system has an absolute meaning. This is why it can be related
to the internal states of its subsystems, as is postulated below.
\vskip0.5cm
{\em Postulate 6. If the reference system $R=I$ is an isolated one 
then the state $\hat \rho_A(I)$ commutes with the
internal state $\hat \rho_A(A)$.}
\vskip0.5cm
This means that the internal state of $A$ coincides with
one of the eigenstates of $\hat \rho_A(I)$.
Note that usually there is no one-to-one correspondence 
between states with respect to different reference systems, thus
one cannot tell (knowing $\hat \rho_A(I)$) which eigenstate 
corresponds to $\hat \rho_A(A)$). In what follows, 
these eigenstates $|\phi_{A,j}>$ will play an important role. 
They will be identified with the corresponding projector 
$\hat \pi_{A,j}=|\phi_{A,j}><\phi_{A,j}|$, and we shall call
them the {\em possible internal states}, as they constitute
the set of those internal states of $A$ that are compatible
with $\hat \rho_A(I)$. (Of course, at a given time only one
of these states exists {\em with respect to $A$}.) For further 
reference, 
we set
\vskip0.5cm
{\em Definition 3. The possible internal states are the eigenstates 
of
$\hat \rho_A(I)$ provided that the reference system $I$ is an
 isolated one.}
\vskip0.5cm
It can happen
that some of the eigenvalues of $\hat \rho_A(I)$ coincides.
Nevertheless, by requiring the continuity of the possible internal 
states as a function of time (using the usual Hilbert space norm)
one can resolve the degeneracy (this degeneracy can remain constantly
 there if $A$ does not interact with other systems,
 but then it is an isolated system and the degenerated 
 eigenvalue is zero, which is irrelevant [cf. {\em Postulate 7}
 below]). Thus the possible internal states 
 are practically unambigously defined once $\hat \rho_A(I)$ 
 is known.
 
 An important property of the possible internal states is given by 
 the following statement:
\vskip0.5cm 
{\em Proposition 1. If $A$ and $B$ are two disjointed physical
 systems (i.e.,
they have no common subsystems) with possible internal states
$|\phi_{A,j}>$ and $|\phi_{B,j}>$, respectively, and the joint
system $A+B$ is an isolated one, then the internal state of
$A+B$ can be written as}
\begin{eqnarray}
|\psi_{A+B}>=\sum_j c_j |\phi_{A,j}>|\phi_{B,j}>\quad.
 \label{e10}
\end{eqnarray}
\vskip0.5cm
Eq.(\ref{e10}) is called the Schmidt canonical form\cite{Schmidt},
for a recent explicit application to simple systems see Ref.\cite{albrecht}. 
It is
straightforward to see that the reverse of {\em Proposition 1}
is also true, i.e., once Eq.(\ref{e10}) holds with some
orthonormed set of states $|\phi_{A,j}>$ and $|\phi_{B,j}>$
($j=1,2,3,..$) then these are the possible internal state of $A$
and $B$, respectively.  As for a proof of {\em Proposition 1}, see 
Appendix A.

\vskip0.5cm
{\em Postulate 7. If $I$ is an isolated system, then the
probability $P(A,j)$ that the
eigenstate $|\phi_{A,j}>$ 
of $\hat \rho_A(I)$ coincides with $\hat \rho_A(A)$ 
is given by the corresponding eigenvalue $\lambda_j$.} 

Due to the
normalization of the internal states these eigenvalues add up to
unity.
\vskip0.5cm
It is important to emphasize that despite of the probability
introduced above, it is not possible to classify reference
systems having the same internal state according to the
internal states of their subsystems, as the former state exists
with respect to the reference system while the latter states
exist only with respect to the subsystems. It is possible, 
however, to learn the internal states of the subsystems 
via suitable measurements (a quantity whose operator 
commutes with $\hat \rho_A(I)$ should be measured), destroying
at the same time the original internal state of $I$. Thus, for
an observer before the measurement the internal state of
$I$ is known and the internal state of $A$ is known afterwards.
Once the former has been recorded, one can speak about the joint
probability of the internal state of $I$ and that of $A$. 
The record, of course,
presupposes that a third system interacts with $I$. 
The probabilities $P(A,j)$ in {\em Postulate 7} are equal to 
those of
the various outcomes of the measurement mentioned above. In
order to establish this connection in terms of the rules of
the present approach two further postulates are needed.
\vskip0.5cm
{\em Postulate 8. The result of a measurement is contained 
unambigously in the internal state of the measuring device.}
\footnote{This statement corresponds to that made in Section 3. 
Note that
in principle it also applies if the 
experience of a 
living being is concerned (cf. Schr\"odinder's cat\cite{cat} or 
Wigner's friend\cite{Wignerf}). Indeed, the present approach does 
not distinguish
between physical systems, they can be either microscopic or
macroscopic, or even living beings. }
\vskip0.5cm
{\em Postulate 9. If $A$ and $B$ are two disjointed physical systems 
with possible internal states
$|\phi_{A,j}>$ and $|\phi_{B,j}>$, respectively, and both systems 
are contained in the isolated reference system $I$, then the joint
probability that $|\phi_{A,j}>$ 
coincides with the internal state of $A$ and at the same time  
$|\phi_{B,k}>$ 
coincides with the internal state of $B$ ($j,k=1,2,3,...$) is given 
by
\begin{eqnarray}
P(A,j,B,k)=Tr_{A+B} \left(\hat \pi_{A,j} \hat \pi_{B,k} \hat 
\rho_{A+B}(I)
\right)\quad,  \label{e16}
\end{eqnarray}
where 

$\hat \pi_{A,j}=|\phi_{A,j}><\phi_{A,j}|,\quad 
\hat \pi_{B,k}=|\phi_{B,k}><\phi_{B,k}|\quad$.}
\vskip0.5cm
As an example let us consider the situation when $A+B$ is an
isolated system. According to {\em Proposition 1} 
(cf. Eq.(\ref{e10}))
$P(A,j,B,k)=|c_j|^2 \delta_{j,k}$. 

Note that it is important that $A$ and $B$ are
disjointed systems (this will play an important role
when explaining why Bell's inequality does not hold, cf. Section
7), as this ensures that the product 
$\hat \pi_{A,j} \hat \pi_{B,k}$ is a single projector acting on the
Hilbert space of $A+B$, which in turn implies (due to the
positive definiteness of $\hat \rho_{A+B}(I)$) that $P(A,j,B,k)$ 
is indeed
real and nonnegative. 

Due to the completeness of the possible internal states
$\sum_k \hat \pi_{A,k}=1$ holds, therefore $\sum_k
P(A,j,B,k)=P(A,j)$, as expected. 

More generally, we can consider the joint probabilities for
more than two disjointed physical systems. Then we set
\vskip0.5cm
{\em Postulate 10. If there are $n$ disjointed physical systems, 
denoted by
\hfill\break
$A_1, A_2, ... A_n$, all contained in the isolated reference 
system $I$ and 
having the 
possible internal states
$|\phi_{A_1,j}>, |\phi_{A_2,j}>,...,|\phi_{A_n,j}>$, respectively, 
then the joint
probability that $|\phi_{A_i,j_i}>$ 
coincides with the internal state of $A_i$ ($i=1,..n$)
is given by
\begin{eqnarray}
P(A_1,j_1,A_2,j_2,...,A_n,j_n)\mbox{\hspace{6cm}}\nonumber\\
=Tr_{A_1+A_2+...+A_n} [\hat \pi_{A_1,j_1} 
\hat \pi_{A_2,j_2}
...\hat \pi_{A_n,j_n}\hat \rho_{A_1+A_2+...+A_n}(I)]\;,\label{e17}
\end{eqnarray}
where $\hat \pi_{A_i,j_i}=|\phi_{A_i,j_i}><\phi_{A_i,j_i}|$.}
\vskip0.5cm
Note that $n=1$ and $n=2$ correspond to {\em Postulate 7} and 
{\em Postulate 9},
respectively.
\vskip0.5cm
Let us discuss now the relation of the probabilities $P(A,j)$
and $P(A,j,B,k)$ with the measurements. The term 'measurement'
means here a usual interaction of the system in question with
another system called the measuring device. For simplicity 
it is assumed that the measurement
process has two special properties (note that this simplification
does not lead to any essential restriction of the generality of 
our discussion): \hfill\break
1. before the measurement the measuring device can be considered
an isolated system\hfill\break
2. if the initial state of the system is a member of a certain
orthonormed set of states $|\varphi_j>$ then the product state 
$|\varphi_j>|m_0>$ evolves during the measurement 
(according to the time dependent Schr\"odinger equation) 
into the product state 
$|\varphi_j>|m_j>$,
where $|m_j>$ ($j=1,2,..$) is an orthonormed set
of states in the Hilbert space of the measuring device.
(Thus we consider {\em quantum nondemolition measurements}
\cite{QNM}.)\hfill\break

As a foregoing general remark let us mention that it is not possible 
to
derive the concept of the above probabilities directly from the 
measurements, 
as the interpretation of the latter's results inevitably needs
the postulates again,
where the probabilities are already involved. What we can show is 
that 
the postulates offer a consistent interpretation of the observed 
relative 
frequences, in the same way as the mathematical probability
theory does, and, moreover, we can reproduce all the usual
results of the quantum mechanics concerning the probabilities.

Let us consider first the case when just $\hat \rho_A(I)$ is
measured ($I$ is isolated), i.e., when the basic dynamics of the
measurement process is given by the relations
\begin{eqnarray}
|\phi_{A,j}>|m_0>\rightarrow |\phi_{A,j}>|m_j>\quad,(j=1,2,..).
\label{e18}
\end{eqnarray}
Therefore, using Eq.(\ref{e10}) with $B=I\setminus A$ the change of 
the 
internal state of the isolated system $I+M$ ($M$ standing for
the measuring device) is given by
\begin{eqnarray}
\left(\sum_j c_j |\phi_{A,j}>|\phi_{B,j}>\right)|m_0>
\rightarrow \sum_j c_j |\phi_{A,j}>|\phi_{B,j}>|m_j>\;.\label{e19}
\end{eqnarray}
Using {\em Postulate 4} and {\em Definition 3} 
one finds that after the measurement the
possible internal states of $A$ are still the $|\phi_{A,j}>$-s,
and the possible internal states of the measuring device are the
$|m_j>$-s. Furthermore, {\em Postulate 7} implies that the
probability to get the $j$-th results (i.e., the probability
that $|m_j>$ coincides with the internal state of $M$, cf. {\em
Postulate 8}) is
$|c_j|^2$, which coincides with the probability $P(A,j)$. 
Using now {\em Postulate 9} we get $P(A,j,M,k)=|c_j|^2
\delta_{j,k}$, i.e., if the $j$-th result is obtained at the 
measurement, then the internal state of $A$ is just
$|\phi_{A,j}>$.

Consider now the joint probability $P(A,j,B,k)$ occuring in 
{\em Postulate 9}.
Suppose that we measure simultaneously $\hat \rho_A(I)$ and 
$\hat \rho_B(I)$ 
by two measuring devices, $M^a$ and $M^b$, respectively. 
Let $C=I\setminus(A+B)$, then the internal state of the system 
$A+B+C+M^a+M^b$
before the measurement is given by
\begin{eqnarray}
\sum_j \sum_k \sum_l c_{j,k,l} |\phi_{A,j}>|m^a_0>
|\phi_{B,k}>|m^b_0>|\phi_{C,l}>\;,
\label{e20}
\end{eqnarray}
and after the measurement it is
\begin{eqnarray}
\sum_j \sum_k \sum_l c_{j,k,l} |\phi_{A,j}>|m^a_j>
|\phi_{B,k}>|m^b_k>|\phi_{C,l}>\;.
\label{e21}
\end{eqnarray}

As a consequence of the orthogonality of the states $|m^a_j>$ one 
concludes
that the possible internal states of $A$ are the $|\phi_{A,j}>$-s, 
and,
due to the orthogonality of the states $|m^b_k>$
that the possible internal states of $B$ are the $|\phi_{B,k}>$-s. 
Therefore, using Eq.(\ref{e16}) we get 
\begin{eqnarray}
P(A,j,B,k)=\sum_l |c_{j,k,l}|^2\quad.
\label{e22}
\end{eqnarray}
 Let us calculate now the possible internal states of the combined
 measuring device $M^a+M^b$. They turn out to be the product states 
 $|m^a_j>|m^b_k>$ ($j,k=1,2,..$), and the probability that they 
 coincide 
 with the internal state of $M^a+M^b$ is
 \begin{eqnarray}
 P(M^a+M^b,(j,k))=\sum_l |c_{j,k,l}|^2
 \label{e23}
\end{eqnarray}
i.e., it coincides with $P(A,j,B,k)$. Using {\em Postulate 10} 
with $n=3$
we can calculate the joint probability $P(A,j,B,k,M^a+M^b,(l,m))$ 
to get
 \begin{eqnarray}
P(A,j,B,k,M^a+M^b,(l,m))=\sum_l
|c_{j,k,l}|^2\delta_{j,l}\delta_{k,m}. \label{e24}
\end{eqnarray}
Thus we can see that if $M^a$ shows the $j$-th result and $M^b$ 
shows the
$k$-th result, then the internal state of $A$ and $B$ has been being 
$|\phi_{A,j}>$ and $|\phi_{B,k}>$, respectively\footnote{Strictly 
speaking, 
in order to see
that the internal states of $A$ and of $B$ have not changed during 
the 
measurement, we need {\em Postulate 11} concerning the time 
evolution, 
see later.}. It is a general
feature that the simultaneous existence of 
internal states of different systems (which also 
means that the reference systems are different) 
 can be defined and provided with probabilities if all they can be 
 brought into correlation with the internal state of the
 same physical system (in the above 
 example this latter has been the combined measuring device), i.e., 
 if they 
 become comparable
 with respect to that physical system. The condition for that is
that the systems in question are disjointed or they are strongly
correlated with disjointed systems. As an example we can
consider the joint probability of the possible internal states
of a system $A$ and that of one of its subsystems $B$. As $B\subset
A$, they are not disjointed, but $A$ is strongly correlated with
$I\setminus A$, where $I$ is an isolated system containing $A$
(cf. {\em Proposition 1}), i.e., if the internal state
of $A$ coincides with the $j$-th possible internal state
$|\phi_{A,j}>$, then the internal state of $I\setminus A$ 
coincides with the corresponding possible internal state
$|\phi_{I\setminus A}>$. The systems $B$ and $I\setminus A$ are
already disjointed. Therefore, $P(A,j,B,k)= P(I\setminus
A,j,B,k)$, the latter probability being already well defined
according to {\em Postulate 9}. Note that it can be brought to
the form 
\begin{eqnarray}
P(A,j,B,k)=<\phi_{B,k}|\hat \rho_B(A,j)|\phi_{B,k}>\quad,\label{e25}
\end{eqnarray}
where 
\begin{eqnarray}
\hat \rho_B(A,j)=Tr_{A\setminus
B}\left(|\phi_{A,j}><\phi_{A,j}|\right)\quad.\label{e25a}
\end{eqnarray}

  Let us outline now how the probability $P(A,j)$ can be related to 
 the actual relative frequences (this procedure is quite similar to 
 that
 used in probability theory). Suppose that we are given $n>>1$ 
 identical isolated
 systems $I_i$ ($i=1,2,...n$), all being in the same internal state, 
 except for a translation, and each containing the identical 
 subsystems $A_i$.
 Let us measure on each $A_i$  the 
 quantity
 $\hat \rho_{A_i}(I)$ ($I=I_1+I_2+...+I_n$) 
 using a measuring device $M_i$, and consider the 
 internal state
 of the combined device $M_1+M_2+...+M_n$. 
  Due to the independence of the measurements we get
 the polynomial distribution for the probability
 that $k_1$ devices show the first result,
 $k_2$ devices the second result etc. This distribution is for $n>>1$ 
 sharply peaked near the mean values 
 $\bar k_1=n P(A,1),\;\bar k_2=n P(A,2),\,...$.
 
 Up to now we discussed rather special measurements in order to
 demonstrate the concepts and their consistency.
 Let us turn now our attention to the question what our approach 
 gives
 in case of an arbitrary measurement. Consider an isolated system 
 $I=A+B$,
 and make a measurement on its subsystem $A$. Suppose that the 
 measurement
 is defined by 
 \begin{eqnarray}
|\varphi_j>|m_0>\rightarrow 
|\varphi_j>|m_j>\quad,
\label{e26}
\end{eqnarray}
where the states $|\varphi_j>$ ($j=1,2,...$) are the eigenstates 
of the 
measured quantity, and therefore constitute a complete orthonormed 
set.
Using Eqs.(\ref{e10}) and (\ref{e26}) the change of the internal 
state of the
isolated system $A+B+M$ during the measurement can be written as
\begin{eqnarray}
\left(\sum_j c_j |\phi_{A,j}>|\phi_{B,j}>\right)|m_0>\nonumber\\
\rightarrow \sum_j \sum_k c_j <\varphi_k|\phi_{A,j}>
|\varphi_k>|m_k>|\phi_{B,j}>\quad.\label{e27}
\end{eqnarray}
Using our postulates, it is easy to show that the possible internal 
states
of $M$ are the $|m_k>$-s with the probability 

$P(M,k)=\sum_j |c_j|^2 |<\varphi_k|\phi_{A,j}>|^2\quad
=<\varphi_k|\hat \rho_A(I)|\varphi_k>$ to coincide with the
 internal state of $M$, which is exactly the
same what we get in the usual formulation of quantum mechanics for
 the 
occurence of the $k$-th result of the measurement.

Let us consider now the time evolution of the states. 
In case of closed systems the usual unitary
time evolution is assumed, i.e.,
\vskip0.5cm
{\em Postulate 11. The internal state $|\psi_C>$ of a closed system 
$C$
satisfies the time dependent Schr\"odinger equation}
\begin{eqnarray}
i\hbar \partial_t |\psi_C>=\hat H |\psi_C> \quad.
\label{e28}
\end{eqnarray}
\vskip0.5cm
Here $\hat H$ stands for the Hamiltonian.

As for the time evolution of the internal states of 
non-closed systems, it can be defined if 
the initial and the final states can be compared, i.e., if the initial
state has been 'recorded' (via suitable interaction with some other
system) in order to make it comparable with the final state (they can 
be compared only if they are related to the internal 
state of the same system at the same time, cf. the discussion of the
joint probabilities above). 

To be more explicit, consider a system $A$ and assume 
that its initial internal state (at time $0$) has been
determined via the measurement of $\hat \rho_A(I)$, where $I$ is
an isolated reference system (cf. Eq.(\ref{e19})).  Suppose now that 
the measuring device $M$ has not interacted with any systems
after this measurement. Actually it is of no importance
whether $M$ interacted with other systems or not, if yes, we can
consider their union with $M$. The important assumption is that
$M$ has not interacted with $A$ after the first measurement. 
Therefore, the internal state $|\psi_M(t)>$ evolves
unitarily according to {\em Postulate 11}, and it is uniquely
related even at a later time $t$ to the initial internal state 
$|\psi_A(0)>$ of $A$. Therefore, the joint probability
$P_t(A,(k|j))$ that
the internal state of $A$ coincides at time $0$ with its $j$-th
possible internal state $|\phi_{A,j}(0)>$ and it coincides at
time $t$ with its $k$-th
possible internal state $|\phi_{A,k}(t)>$ is given by 
\begin{eqnarray}
P_t(A,(k|j))=P(M,j,A,k)(t)\quad,
\label{e29}
\end{eqnarray}
thus, it is expressed through a joint probability of type 
(\ref{e16}).

\section{Resolution of Schr\"odinger's cat paradox}

\subsection{The paradox within the framework of the Copenhagen interpretation}
The well known thought experiment contains a radioactive nucleus, whose
decay triggers through a detector a device which breaks a poison capsule,
letting some poisonous gas out, and thus killing a cat. At a given instant of time the state of the nucleus is a superposition of a decayed and a nondecayed state, 
therefore, due to the linearity of the time dependent Schr\"odinger equation
the state of the whole system is given by
\begin{eqnarray}
\alpha |+>|d_+>|d>+\beta|->|d_->|a>\;,\label{cat}
\end{eqnarray}
where $|+>$ and $|->$ stand for the decayed and nondecayed state of the
nucleus, respectively, $|d_+>$ and $|d_->$ describe the corresponding states
of the detector+device+poison system, finally, $|d>$ and $|a>$ stand
for the dead and the alive state of the cat, respectively.
If one looks at the cat, according to the Copenhagen interpretation
the collapse of the wave function takes place, and one finds
either that the cat is dead or that it is living. 
Before looking at it, however, Eq.(\ref{cat}) should hold, which
contains both the living and the dead states, so one should
conclude that the cat is neither living nor dead, until someone
looks at it. This unphysical conclusion is the paradox. We formulate it now
in somewhat different form. Physically we expect that the cat is either
living or dead. Therefore, even if noone looks at it, its state
must be either $|a>$ or $|d>$. In contrast, if one calculates
the state of the cat, i.e., its reduced density matrix,
it is
\begin{eqnarray}
\hat \rho_c=|d>|\alpha|^2<d|+|a>|\beta|^2<a|\;,
\end{eqnarray}
which contains both states. One may make the contradiction even
sharper, if an observer is also included (here we assume the universal
applicability of the quantum mechanical description). Denoting the state
of the observer by $|o_d>$ if he finds the cat dead and by $|o_a>$
if he finds it alive, the state of the whole system will be
\begin{eqnarray}
\alpha |+>|d_+>|d>|o_d>+\beta|->|d_->|a>|o_a>\;,\label{cato}
\end{eqnarray}
and the state of the observer is
\begin{eqnarray}
\hat \rho_o=|o_d>|\alpha|^2<o_d|+|o_a>|\beta|^2<o_a|\;.\label{obs}
\end{eqnarray}
There is no doubt that in this situation the observer would 
experience something definite, i.e., its state must be either
$|o_a>$ or $|o_d>$. This contradicts Eq.(\ref{obs}), therefore
one may say that the quantum mechanical prediction in this situation
contradicts the experience.

\subsection{Resolution of the paradox according to the present approach}

Our approach specifies what corresponds to the experience of the cat or
that of
the observer (cf. {\em Postulate 8} and the footnote there): it is the
internal state of the cat or that of the observer. Applying Postulate 6, 
the internal state of the cat can be $|a>$ or $|d>$, and the internal state
of the observer can be $|o_a>$ or $|o_d>$. This theoretical
prediction is in complete
agreement with the actual experience, therefore, according to the
present approach no paradox appears. 

Note that {\em Postulate 9} implies that 
if the cat's internal state is $|d>$ ($|a>$), then
the observer's internal state is $|o_d>$ ($|o_a>$)
as physically expected.

\section{Resolution of the EPR paradox}

\subsection{A review of the paradox within the framework of the Copenhagen
interpretation}
 
Let us review briefly the EPR paradox first within the framework 
of the 
Copenhagen interpretation \cite{EPR}. 
For simplicity, the paradox will be discussed in the
case of a spin system\cite{bohm}.
Consider a system consisting of two different spin-half fermions,
which have interacted in the past but which do not interact
any longer. Let us concentrate on the spin degrees of freedom.
Suppose that the initial state of this two particle system is
given by
\begin{eqnarray}
a|1,\uparrow>|2,\downarrow>-b|1,
\downarrow>|2,\uparrow>\quad,
\label{e30}
\end{eqnarray}
where $a$ and $b$ are complex numbers, satisfying
\begin{eqnarray}
|a|^2+|b|^2=1\quad.
\label{e31}
\end{eqnarray}
The notation $|1,\uparrow>$ stands for such a state of the first
particle, where the $z$ component of the spin (denoted by
$\hat S_{1z}$) has the definite value $+\frac{1}{2}$. The other
notations have an analogous meaning. Consider now what happens
if one performs a measurement on the first particle. Let us 
consider the situation
when one measures $\hat S_{1z'}$, where the $z'$ axis is
obtained from the $z$ axis by
a rotation at an angle $\delta$ around the $x$ axis. Note
that in this section all the rotations will be given relative to
the first (unprimed) coordinate system. This coordinate system
will be called the original one.
The initial state of the whole
system (including the measuring device) is given by
\begin{eqnarray}
|m_0>  (
a|1,\uparrow>  |2,\downarrow>-b|1,
\downarrow>  |2,\uparrow>)\quad,
\label{e32}
\end{eqnarray}
where $|m_0>$ stands for the initial state of the measuring device.
The time evolution during the measurement 
can be established by using the relations
\begin{eqnarray}
|m_0>  |1,\delta,\uparrow>\rightarrow |m_+>
  |1,\delta,\uparrow>
  \label{e33}
\end{eqnarray}
and
\begin{eqnarray}
|m_0>  |1,\delta,\downarrow>\rightarrow |m_->
  |1,\delta,\downarrow>\quad,
  \label{e34}
\end{eqnarray}
where
\begin{eqnarray}
|1,\delta,\uparrow>=\cos\left(\frac{\delta}{2}\right)|1,\uparrow>
-\sin\left(\frac{\delta}{2}\right)|1,\downarrow>\quad,
\label{e35}
\end{eqnarray}
and
\begin{eqnarray}
|1,\delta,\downarrow>=\sin\left(\frac{\delta}{2}\right)|1,\uparrow>
+\cos\left(\frac{\delta}{2}\right)|1,\downarrow>\quad.
\label{e36}
\end{eqnarray}
Thus the final state of the whole system is
\begin{eqnarray}
|m_+> |1,\delta,\uparrow> 
\left(a\,\cos\left(\frac{\delta}{2}\right)|2,\downarrow>
+b\,\sin\left(\frac{\delta}{2}\right)|2,\uparrow>\right)\nonumber\\
+|m_-> |1,\delta,\downarrow> 
\left(a\,\sin\left(\frac{\delta}{2}\right)|2,\downarrow>
-b\,\cos\left(\frac{\delta}{2}\right)|2,\uparrow>\right)\;.
\label{e37}
\end{eqnarray}
According to the Copenhagen interpretation, the wave function
collapses and thus, if $S_{1,z'}=+\frac{1}{2}$ has been
measured, the second particle will have the wave function
\begin{eqnarray}
\left(|a|^2\cos^2\left(\frac{\delta}{2}\right)+|b|^2\sin^2\left(\frac{\delta}{2}\right)\right)^
{-\frac{1}{2}}
\left(a\,\cos\left(\frac{\delta}{2}\right)|2,\downarrow>
+b\,\sin\left(\frac{\delta}{2}\right)|2,\uparrow>\right)
\label{e38}
\end{eqnarray}
which is the eigenfunction of the $\hat S_{2,z''}$ spin component
corresponding to the eigenvalue $-\frac{1}{2}$. Here $z''$
stands for the $z$ axis of a new coordinate system obtained from
the original one by a rotation with the
Eulerian angles $\alpha,
\beta,\gamma$ given by
\begin{eqnarray}
\exp(i\alpha)=\frac{ab}{|ab|}\quad,
\label{e39}
\end{eqnarray}
\begin{eqnarray}
\tan\left(\frac{\beta}{2}\right)=\frac{|b|}{|a|}\tan\left(\frac{\delta}{2}\right)\quad,
\label{e40}
\end{eqnarray}
\begin{eqnarray}
\exp(-i\gamma)=\frac{b/a}{|b/a|}\quad.
\label{e41}
\end{eqnarray}
Similarly, if $S_{1,z'}=-\frac{1}{2}$ has been measured, the wave 
function of
the second particle will be
\begin{eqnarray}
\left(|a|^2\sin^2\left(\frac{\delta}{2}\right)
+|b|^2\cos^2\left(\frac{\delta}{2}\right)\right)^
{-\frac{1}{2}}
\left(a\,\sin\left(\frac{\delta}{2}\right)|2,\downarrow>
-b\,\cos\left(\frac{\delta}{2}\right)|2,\uparrow>\right)
\label{e42}
\end{eqnarray}
which is the eigenfunction of the $\hat S_{2,z'''}$ spin component
corresponding to the eigenvalue $+\frac{1}{2}$. Here $z'''$
stands for the $z$ axis of another coordinate system obtained
from the original one by a rotation with the
Eulerian angles $\alpha',
\beta',\gamma'$ given by
\begin{eqnarray}
\exp(-i\alpha')=\frac{ab}{|ab|}\quad,
\label{e43}
\end{eqnarray}
\begin{eqnarray}
\tan\left(\frac{\beta'}{2}\right)=\frac{|a|}{|b|}\tan\left(\frac{\delta}{2}\right)\quad,
\label{e44}
\end{eqnarray}
\begin{eqnarray}
\exp(i\gamma')=\frac{a/b}{|a/b|}\quad.
\label{e45}
\end{eqnarray}
Summing up, having performed the measurement of $\hat S_{1,z'}$,
if the result is
$S_{1,z'}=+\frac{1}{2}$,
one can predict for sure that
$S_{2,z''}=-\frac{1}{2}$, and if the result is
$S_{1,z'}=-\frac{1}{2}$, then
one can predict for sure that
$S_{2,z'''}=+\frac{1}{2}$. Obviously, the second particle has not
been disturbed in any way. According to the EPR argument, if the 
value of 
a quantity 
can be predicted without disturbing the corresponding system, then 
there
must exist an element of the physical reality that corresponds to 
this 
quantity. 

Therefore, either $S_{2,z''}$ or
$S_{2,z'''}$ is an element of the reality. But these quantities 
depend on the
measurement done on the first particle, as they depend on $\delta$.
One expects that\hfill\break
{\em STATEMENT I. the elements of the reality attached to the
second particle are independent of what have happened with
the first particle, as they are separated and do not
interact.}\hfill\break
Therefore, it does not matter that one can perform on the
first particle only one measurement at the same time 
(cf. the next to last paragraph in \cite{EPR}). 
Either $S_{2,z''}$ or
$S_{2,z'''}$ must be an element of the physical reality 
for any $\delta$-value.

We may set different values for $\delta$, say \hfill\break
a, $\delta=0$ \hfill\break
 and \hfill\break
b, $\delta=\vartheta\ne 0$. \hfill\break
In the first case, $z''$ and $z'''$ coincide with the $z$ 
axis.
 Thus we see that either $S_{2,z}$ and $S_{2,z''}$ or
$S_{2,z}$ and $S_{2,z'''}$ are elements of the reality at the same
time (here $z''$ and $z'''$ correspond to $\delta=\vartheta$).

On the other hand, $\hat S_{2,z}$ commutes neither with
$\hat S_{2,z''}$, nor with $\hat S_{2,z'''}$ (except for
some special values of $\vartheta$ that are not of interest now)
that implies that these quantities can be elements of the
 reality at the same time only if \hfill\break
       {\em STATEMENT II. the quantum mechanical description is incomplete.}
       \hfill\break (cf. the alternatives (1) and (2) in \cite{EPR}.)
As has been shown, {\em STATEMENT I.} implies {\em STATEMENT II.}.       
 
The first statement is usually identified with the requirement
of the locality. Therefore, Einstein, Podolsky and Rosen accept
that it is true, thus they conclude that {\em STATEMENT II.} is also true,
i.e., the quantum mechanical description is incomplete.
 
Sometimes people reject {\em STATEMENT II.} and therefore reject
{\em STATEMENT I.} as well,
 saying
that according to quantum mechanics, Nature is nonlocal\cite{Esp}.

\subsection{Solution of the paradox according to the present approach}

Let us consider now what the
present approach says. This time we have to consider
an isolated system (otherwise most of our postulates do not work),
therefore we include the preparation process which prepares the state
(\ref{e30}). The macroscopic preparation device $M_p$ and the two particle
system $P_1+P_2$ together are assumed to constitute initially an isolated 
system\footnote{We may include in $M_p$ even that part of the
environment which
interacts with the actual preparation device.}. 
The initial internal state of the system $M_p+P_1+P_2$ evolves
unitarily (according to the time dependent Schr\"odinger equation) to
an entangled state which can be written, using the Schmidt canonical form
(\ref{e10}) as
\begin{eqnarray}
|\mu,1,2>=\sum_j c_j |\mu_j>|\psi^\bot_j(1,2)>+\alpha |\mu_p>|\psi_p>
\end{eqnarray}
Here $|\psi_p>$ stands for the prepared state (\ref{e30}), the states
$|\psi^\bot_j(1,2)>$ are orthogonal to it and together constitute an 
orthonormed set of states. Note that the preparation device is
constructed such a manner (mathematically, its Hamiltonian has such a form)
that the state $|\psi_p>$ will be practically the same, while
all the other states and coefficients may vary from one preparation process
to the other, corresponding to the different initial conditions. 
The states $|\mu_j>$, $|\mu_p>$ 
of the preparation device also
constitute an orthonormed set of states. It is important to note that
the coordinate dependence of the state $|\psi_p>$ (that is suppressed
in Eq. (\ref{e30})) is such that it describes the two particle system
entering the apparatus where the measurement will be done. In contrast,
the states $|\psi^\bot_j(1,2)>$ describe the two particle system
when it does not enter the apparatus. This implies mathematically that the
dynamics of the measurement done on the first particle
(cf. Eqs.(\ref{e33}), (\ref{e34})) should be
completed by the rule 
\begin{eqnarray}
|m_0>|\psi^\bot_j(1,2)>\rightarrow |m_0>|\psi^\bot_j(1,2)>\;.
\end{eqnarray}
Therefore, the state of the isolated system $M_p+P_1+P_2+M$ 
after the measurement can be written as
\begin{eqnarray}
|\mu,1,2,m>=|m_0>\left(\sum_j c_j |\mu_j>|\psi^\bot_j(1,2)>\right)\nonumber\\+
\alpha\left[|m_+>|\mu_p> |1,\delta,\uparrow> 
\left(a\,\cos\left(\frac{\delta}{2}\right)|2,\downarrow>
+b\,\sin\left(\frac{\delta}{2}\right)|2,\uparrow>\right)\right.\\
\left.+|m_-> |\mu_p>|1,\delta,\downarrow> 
\left(a\,\sin\left(\frac{\delta}{2}\right)|2,\downarrow>
-b\,\cos\left(\frac{\delta}{2}\right)|2,\uparrow>\right)\right]\;.\nonumber
\label{vegall}
\end{eqnarray}
According to the 
rules given in the previous Section (see {\em Postulate 4} and
 {\em Definition 3}), the possible internal states of the two
particle system are 
\begin{eqnarray}
|\psi^\bot_j(1,2)>\;,\label{noprep}
\end{eqnarray}
\begin{eqnarray}\left(|a|^2\cos^2\left(\frac{\delta}{2}\right)
+|b|^2\sin^2\left(\frac{\delta}{2}\right)\right)^
{-\frac{1}{2}}\mbox{\hspace{3cm}}\nonumber\\
\times |1,\delta,\uparrow> 
\left(a\,\cos\left(\frac{\delta}{2}\right)|2,\downarrow>
+b\,\sin\left(\frac{\delta}{2}\right)|2,\uparrow>\right)
\label{e46}
\end{eqnarray}
and
\begin{eqnarray}\left(|a|^2\sin^2\left(\frac{\delta}{2}\right)
+|b|^2\cos^2\left(\frac{\delta}{2}\right)\right)^
{-\frac{1}{2}}\mbox{\hspace{3.5cm}}\nonumber\\
\times |1,\delta,\downarrow> 
\left(a\,\sin\left(\frac{\delta}{2}\right)|2,\downarrow>
-b\,\cos\left(\frac{\delta}{2}\right)|2,\uparrow>\right)\quad.
\label{e47}
\end{eqnarray}
The states (\ref{noprep}) correspond to the case when the preparation
was not successful and the first particle did not interact with the
measuring device $M$. Certainly, this case is of no interest for us, and
we shall concentrate on the states
(\ref{e46}), (\ref{e47}).
Choosing the reference system $R$ to be the two 
particle system 
$P_1+P_2$, we get that the state of the second particle with 
respect to the
two particle system is 
\begin{eqnarray}
\hat \rho_{P_2}(P_1+P_2)=|\xi><\xi|\quad,
\label{e48}
\end{eqnarray}
where
\begin{eqnarray}
|\xi>=\left(|a|^2\cos^2\left(\frac{\delta}{2}\right)
+|b|^2\sin^2\left(\frac{\delta}{2}\right)\right)^
{-\frac{1}{2}}\nonumber\\\times
\left(a\,\cos\left(\frac{\delta}{2}\right)|2,\downarrow>
+b\,\sin\left(\frac{\delta}{2}\right)|2,\uparrow>\right)\;,
\label{e49}
\end{eqnarray}
if the internal state of the two particle system coincides with 
Eq.(\ref{e46}) and it is 
\begin{eqnarray}
|\xi>=\left(|a|^2\sin^2\left(\frac{\delta}{2}\right)
+|b|^2\cos^2\left(\frac{\delta}{2}\right)\right)^
{-\frac{1}{2}}\nonumber\\\times
\left(a\,\sin\left(\frac{\delta}{2}\right)|2,\downarrow>
-b\,\cos\left(\frac{\delta}{2}\right)|2,\uparrow>\right)\;,
\label{e50}
\end{eqnarray}
if the internal state of the two particle system coincides with 
Eq.(\ref{e47}).
Comparing Eqs.(\ref{e49}), (\ref{e50}) with Eqs.(\ref{e38}), 
(\ref{e42}), 
respectively, we see that the state 
of the second particle which was derived using the assumption of 
the collapse
of the wave function is actually $\hat \rho_{P_2}(P_1+P_2)$. For
a more general derivation of this result see Appendix B. 

In the present approach the states 
(which depend on the quantum reference systems)
represent the elements of the reality\footnote{Note 
that from this point of view the EPR criterion
concerning the elements of the reality is only a sufficient
one.}. The correspondence 
with the earlier discussion is clear: $S_{2,z''}$ is said to be 
an element of the
reality if $|\xi>$ (cf. Eq.(\ref{e48})) is an eigenstate of $\hat
S_{2,z''}$. The paradox can be formulated now
that if one rejects {\em STATEMENT II.}, then one has to reject 
{\em STATEMENT I.} as well,
i.e., the state $\hat \rho_{P_2}(P_1+P_2)$ 
representing an element of the reality 
attached to the second particle depends on the measurement 
done on the first particle.
But this state describes $P_2$ with respect 
to the quantum reference system $P_1+P_2$.
 Therefore we conclude as follows: 

{\em The dependence of the state $\hat \rho_{P_2}(P_1+P_2)$ 
on the measurement 
done on the first particle is attributed now to the disturbance 
of the quantum reference system $P_1+P_2$
rather than to any nonlocal influence on the second particle.}

Hence, one can see that in the present approach {\em STATEMENT I.} is not
necessarily implied by locality. 
What locality actually implies is that the 
internal state $\hat \rho_{P_2}(P_2)$ of the second particle 
       must        be independent 
        of the measurement done on the first particle, 
as in this case neither the system
to be described, 
nor the quantum reference system is influenced by the measurement.\newpage

 Let us show now
that it is fulfilled. (As we have already postulated the
unitary time evolution of the internal states of closed systems,
here we actually check the consistency of the postulates.) The 
arguments become more transparent if we consider more generally
the question whether the internal state of a system $A=P_2$ is
influenced by the interaction of another, separated system $B=P_1$
with a further separated system $C=M$, assuming that $A$ and $B$
have interacted before, while $A$ and $C$ not. The composed
system $A+B+C$ is assumed to be an isolated one. Under these
assumptions the internal state of the system $A+B+C$ can 
be written as
\begin{eqnarray}
|\psi_{A+B+C}>(0)
=\left(\sum_j c_j |\phi_{A,j}>|\phi_{B,j}>\right)|\psi_C>.
\label{e51}
\end{eqnarray}
As before, $|\phi_{A,j}>$ and $|\phi_{B,j}>$ stand for the
possible internal states of $A$ and of $B$, respectively, while
$|\psi_C>$ is the internal state of $C$ (it is still isolated).
According to the assumptions, the time evolution operator $\hat
U_t$ factorizes as
\begin{eqnarray}
\hat U_t=\hat U_t(A) \hat U_t(B+C)\quad,
\label{e52}
\end{eqnarray}
where the arguments correspond to the systems whose Hilbert
space is acted upon by the operators. Therefore,
\begin{eqnarray}
|\psi_{A+B+C}>(t)
=\sum_j c_j \left(\hat U_t(A)|\phi_{A,j}>\right)
\left( \hat U_t(B+C)|\phi_{B,j}>|\psi_C>\right)\;.
\label{e53}
\end{eqnarray}
As $\hat U_t(A)|\phi_{A,j}>$  and 
$\hat U_t(B+C)|\phi_{B,j}>|\phi_C>$ ($j=1,2,...$) constitute
orthonormed set of states in the Hilbert space of $A$ and of
$B+C$, respectively \footnote{For $t=0$ it follows by
assumption, and for later times it is due to the unitarity of
the time evolution operators $\hat U_t(A),\hat U_t(B+C)$.},
they are the possible internal states of $A$ and of $B+C$,
respectively, according to the reverse of {\em Proposition 1}. 
Thus we see that the possible internal states of
$A$ are independent of the interaction of $B$ and $C$. According
to {\em Postulate 11} the actual internal state of $A$ is
independent of it, too. We get the same result if we apply 
Eq.(\ref{e29}) 
with the replacement $M=B+C$. Hence, we can say that
quantum mechanics fulfilles any reasonable requirement of
locality. 

\section{Violation of Bell's inequality}

We discuss now the question of the
violation of Bell's inequality\cite{hid}. 
As a brief reminder, let us think again of two spin-half particles 
in an entangled state, 
which is due to some previous interaction between them.
Suppose that the particles are already separated so much that
they can no longer interact with each other. 
Imagine that we perform spin measurements in different directions
 on each
of the two separated particles. 
It is natural to assume that any
correlation between the results of the measurements performed
on the different particles can come only from the previous
interaction which created the entangled state. One may also assume that
there are some
stable properties attached to each system, so that these
properties 'store' the correlation after the systems have become
separated. Using these assumptions one may derive that the
 correlations cannot be
arbitrary but must satisfy a certain inequality, namely
\footnote{Here $P(\alpha +,\beta +)$ stands for the
probability that measuring the spin of the first particle
along a direction at an angle $\alpha$ compared to the $z$ axis,
the result is $+\frac{1}{2}$ and at the same time measuring 
the spin of the second particle
along a direction at an angle $\beta$ compared to the $z$ axis,
the result is $+\frac{1}{2}$.}
\begin{eqnarray}
P(\alpha +,\beta +)
\le P(\alpha +,\gamma +)
+P(\gamma +,\beta +)\quad.\label{bell}
\end{eqnarray}
 This is Bell's
inequality. The correlations may be calculated quantum
mechanically, and the quantum prediction {\em does not} always 
satisfy Bell's inequality. Correlations are measurable
quantities, and experiments\cite{exp} have proved the correctness of 
the  quantum prediction and thus the violation of Bell's
inequality.

Most people seem to believe that the above
result implies that separated systems can influence each
other ('nonlocality'). This belief is based on the careful analysis of the
above sketched derivation of Bell's inequality. It turns out that
this derivation
is completely independent of quantum mechanics, and it is
based on a few very fundamental assumptions\cite{SciAm}, \cite{Esp}: realism,
inductive inference and Einstein separability. Realism and
inductive inference are so important in physics that we certainly 
do not want 
to give up them. The usual conclusion is that
Einstein separability is violated.

Nevertheless, we maintain that such a conclusion is 
physically unacceptable. The principle of Einstein
separability has served us well in every branch of physics,
even in quantum physics (apart from the measurement process), 
including the most sophisticated
quantum field theories. 
The only way out can be if there is some further,
independent and hidden assumption, which seems to us 
obvious, but which is not valid in quantum mechanics. Then the
violation of Bell's inequality implies the invalidity
of this assumption rather than that of Einstein separability or locality.

In this section we show that it is indeed the
case. The hidden, not allowed assumption mentioned above
is connected to the fact that in the present theory 
it may happen that different states (corresponding to different quantum
reference systems), although individually exist,
cannot be compared (cf. Section 3).  
In case of the violation of Bell's inequality it turns out that
the states of the measuring 
devices and those which 'store' the correlations are not
comparable as any attempt for a comparison changes the correlations themselves.
Therefore, the usual picture about stable properties 
which 'store' the correlations and are comparable in principle 
at any time with anything
does not apply,
although the correlations may be attributed exclusively
to the 'common past' (previous interaction) of the particles.

We discuss now quantitatively the situation mentioned above,
i.e., when measurements on both
particles are performed. 
Before the measurements the internal
state of the isolated system $P_1+M_1+P_2+M_2$ ($P_1,P_2$
standing for the particles and $M_1,M_2$ for the measuring
devices, respectively) is given by
\begin{eqnarray}
\left(\sum_j c_j
|\phi_{P_1,j}>|\phi_{P_2,j}>\right)|m^{(1)}_0> |m^{(2)}_0> \quad,
\label{e54}
\end{eqnarray}
while it is
\begin{eqnarray}
\sum_j c_j
\hat U_t(P_1+M_1)\left(|\phi_{P_1,j}>|m^{(1)}_0>\right)
\hat U_t(P_2+M_2)\left(|\phi_{P_2,j}>|m^{(2)}_0>\right) \quad,
\label{e55}
\end{eqnarray}
with a time $t$ later, i.e. during and after the measurements. 
In Eqs.(\ref{e54}), (\ref{e55}) 
the general notation have been used in order to exhibit
 the mathematical structure. Evidently,
$j$ can take on the value $1$ and $2$, 
$c_1=a$, $c_2=-b$ ($|a|^2+|b|^2=1$) and
\begin{eqnarray}
|\phi_{P_1,1}>=|1,\uparrow>\nonumber\\
|\phi_{P_1,2}>=|1,\downarrow>\nonumber\\
|\phi_{P_2,1}>=|2,\downarrow>\label{e56}\\
|\phi_{P_2,2}>=|2,\uparrow>.\nonumber
\end{eqnarray}

According to Eq.(\ref{e55}) $P_1+M_1$ and $P_2+M_2$ are closed systems,
their internal states evolve unitarily and do not influence each
other. This time evolution can be derived from the 
relations
\begin{eqnarray}
|\xi(P_i,j)>|m^{(i)}_0>\rightarrow |\xi(P_i,j)>|m^{(i)}_j>\quad,
\label{e58}
\end{eqnarray}
where $i,j=1,2$ and $|\xi(P_i,j)>$ is the $j$-th eigenstate of the 
spin measured 
on the $i$-th particle along an axis $z^{(i)}$ which closes an angle
$\vartheta_i$ with the original $z$ direction. Denoting the 
internal state of $P_i$ by $|\psi_{P_i}>$, we get
\begin{eqnarray}
|\psi_{P_i}>|m_0^{(i)}>\;\rightarrow \;
\sum_j <\xi(P_i,j)|\psi_{P_i}>|\xi(P_i,j)>|m_j^{(i)}>\;.\label{e58a}
\end{eqnarray}
As we see, the $i$-th measurement process 
is completely determined by the initial internal state of the
particle $P_i$. Therefore, any correlation between
the measurements may only stem from the initial correlation
of the internal states of the particles.

Using Eq.(\ref{e58}) the final state (\ref{e55}) may be written as
\begin{eqnarray}
\sum_{j,k}\left(
\sum_l c_l<\xi(P_1,j)|\phi_{P_1,l}><\xi(P_2,k)|\phi_{P_2,l}>
\right)\nonumber\\
\times |m^{(1)}_j>|m^{(2)}_k>|\xi(P_1,j)>|\xi(P_2,k)>\;.\label{finst1}
\end{eqnarray}
According to {\em Postulate 6} and {\em Definition 3} the
possible internal states of $M_1$ and $M_2$ are the
$|m^{(1)}_j>$-s and the $|m^{(2)}_k>$-s, respectively. 

The probability to observe the $j$-th result (up or
down spin in a chosen direction) in the $i$-th measurement ($i=1,\,2$) is
$
P(M_i,j)=\sum_l |c_l|^2 |<\xi(P_i,j)|\phi_{P_i,l}>|^2$.
This may be interpreted in conventional terms: $|c_l|^2$ is
the probability that $|\psi_{P_i}>=|\phi_{P_i,l}>$,
and $|<\xi(P_i,j)|\phi_{P_i,l}>|^2$ is the conditional
probability that one gets the $j$-th result if $|\psi_{P_i}>=|\phi_{P_i,l}>$.
Thus we see that the initial internal state of $P_i$ determines
the outcome of the $i$-th measurement in the usual 
probabilistic sense.  
But doesn't it mean that the internal states of $P_1$ and $P_2$
play the role of local hidden variables? No, because hidden variables
are thought to be comparable with the results of the measurements
so that their joint probability may be defined, while in our theory
there is no way to define the joint probability
$P(P_1,l_1,P_2,l_2,(0);M_1,j,M_2,k,(t))$, i.e., the probability that initially
$|\psi_{P_1}>=|\phi_{P_1,l_1}>$ and $|\psi_{P_2}>=|\phi_{P_2,l_2}>$
{\em and} finally $|\psi_{M_1}>=|m^{(1)}_j>$ 
and $|\psi_{M_2}>=|m^{(2)}_k>$. Intuitively we would write
\begin{eqnarray}
P(P_1,l_1,P_2,l_2,(0);M_1,j,M_2,k,(t))\mbox{\hspace{6cm}}\nonumber\\
=|c_{l_1}|^2\delta_{l_1,l_2}|<\xi(P_1,j)|\phi_{P_1,l_1}>|^2
|<\xi(P_2,k)|\phi_{P_2,l_2}>|^2,\mbox{\hspace{2.3cm}}\label{u18}
\end{eqnarray}
as $|c_{l_1}|^2\delta_{l_1,l_2}$ is
the joint probability that $|\psi_{P_1}>=|\phi_{P_1,l}>$ 
and $|\psi_{P_2}>=|\phi_{P_2,l}>$, and $|<\xi(P_i,j)|\phi_{P_i,l_i}>|^2$ is the conditional
probability that one gets the $j$-th result in the $i$-th
measurement if initially $|\psi_{P_i}>=|\phi_{P_i,l_i}>$ ($i=1,2$).
Certainly the existence of such a joint probability would immediately imply the
validity of Bell's inequality, thus it is absolutely important 
to understand why this probability does not exist.
Here we mention that the so called modal interpretation\cite{modal}, which is equivalent
 with assuming the existence of the internal state of the subsystem in an absolute sense
 (i.e., independently of quantum reference systems) would lead to Bell's inequality.
 Indeed, the absolute existence of the the initial internal states of $P_1$ and of $P_2$ 
 would mean that initially the particle pairs can be conceived as if they were
 sorted according to the internal state of $P_1$ and of $P_2$, and in each case
 one can consider the measurements as done independently (due to their separation) 
 on the corresponding internal state, that implies Eq.(\ref{u18}).

Returning to our approach, let us remark, first of all, 
that using  {\em  Postulate 9}, we may calculate the correlation
between the measurements, i.e., the joint probability 
that $|\psi_{M_1}>=|m^{(1)}_j>$ {\em and} $|\psi_{M_2}>=|m^{(2)}_k>$.
We obtain
\begin{eqnarray}
P(M_1,j,M_2,k)
=\left|\sum_l c_l<\xi(P_1,j)|\phi_{P_1,l}><\xi(P_2,k)|\phi_{P_2,l}>\right|^2
\;.\label{u19}
\end{eqnarray}
This is the usual quantum mechanical expression 
which violates Bell's inequality and whose correctness is experimentally
proven. Thus our theory gives the correct expression for the correlation.
Nevertheless, if the joint probability 
(\ref{u18}) exists, it leads to 
\begin{eqnarray}
P(M_1,j,M_2,k)
=\sum_l |c_l|^2 |<\xi(P_1,j)|\phi_{P_1,l}>|^2
|<\xi(P_2,k)|\phi_{P_2,l}>|^2\label{u20}
\end{eqnarray}
which satisfies Bell's inequality and contradicts
Eq.(\ref{u19}). 

Let us demonstrate that no such contradiction appears.
Evidently, the joint probability 
$P(P_1,l_1,P_2,l_2,(0);M_1,j,M_2,k,(t))$ can be physically meaningful 
only if one can compare the initial internal states of $P_1$ and $P_2$
with the final internal states of $M_1$ and $M_2$ by suitable
nondisturbing measurements. 
If we try to compare the initial internal 
states of $P_1$ and of $P_2$ with the final
internal states of $M_1$ and $M_2$, 
the first difficulty appears because we want to compare states
given at different times. Nevertheless, as the initial internal state
of $P_i$ is uniquely related to the final internal state of the
system $P_i+M_i$, the joint probability $P(P_1,l_1,P_2,l_2,(0);M_1,j,M_2,k,(t))$ 
(if exists) coincides with $P(P_1+M_1,l_1,P_2+M_2,l_2,M_1,j,M_2,k)$,
where all the occuring states are given after the measurements.
As the systems $P_1+M_1,\;P_2+M_2,\;M_1,\;M_2$ are not disjointed, 
our {\em Postulates} do not provide us with an expression for
the joint probability we are seeking for. 
 If we check $|\psi_{M_1}>$ and $|\psi_{M_2}>$ by
 suitable nondisturbing measurements, we destroy
 $|\psi_{P_1+M_1}>$ and $|\psi_{P_2+M_2}>$, inhibiting any comparison.
  On the other hand, if we check
 first $|\psi_{P_1+M_1}>$ and $|\psi_{P_2+M_2}>$, then 
$P(M_1,j,M_2,k)$ changes.
In fact, after suitable measurements 
performed on
$P_i+M_i$ (which do not change the internal states of $P_i+M_i$)
 by further measuring devices $\tilde M_i$ we get for the
internal state  of the whole system
 \begin{eqnarray}
\sum_l c_l\left( \sum_j <\xi(P_1,j)|\phi_{P_1,l}>|\xi(P_1,j)>|m^{(1)}_j>\right)
\mbox{\hspace{2cm}}\nonumber\\
\times\left( \sum_k <\xi(P_2,k)|\phi_{P_2,l}>|\xi(P_2,k)>|m^{(2)}_k>
\right)
|\tilde m^{(1)}_l>|\tilde m^{(2)}_l>\;.\label{u21}
\end{eqnarray} 
This is exactly the same expression as that we would get if the initial 
internal states of $P_1$ and $P_2$ were recorded by $\tilde M_1$ 
and $\tilde M_2$, respectively.
As the systems $M_1,\;M_2,\;\tilde M_1,\;\tilde M_2$ are disjointed,
we may apply {\em Postulate 10} for $n=4$ and we indeed get for 
$P(\tilde M_1,l_1,\tilde M_2,l_2,M_1,j,M_2,k)$ the expression
(\ref{u18}). Do we get then a contradiction with Eq.(\ref{u19})?
No, because applying {\em Postulate 9} directly, we get in this case
Eq.(\ref{u20}) instead of Eq.(\ref{u19}). Thus we see that the
extra measurements have changed the correlations and our theory 
gives account of this effect consistently.

Note that the extra measurements 
have not changed the internal state of $P_1$ and of
$P_2$, nor have they modified the time evolution 
of the internal state of $P_1+M_1$ 
and of $P_2+M_2$ during the other two measurements. They have, however,
changed the correlation between the measurements. It may seem
misterious how this can be, once the measurements themselves have not changed.
The explanation is again connected 
to the dependence of the states on the quantum reference systems.
Considering, say, the time evolution of the internal state of $P_1+M_1$ 
during the first measurement (using $M_1$), the quantum reference system 
is $P_1+M_1$, whose internal state has been the same initially as before.
The probability distribution 
$P(M_1,j)=\sum_k P(M_1,j,M_2,k)$ is unchanged, too
(cf. Eqs.(\ref{u19}), (\ref{u20})). 
If one wants to see the correlations,
the time evolution of the internal state of the system $P_1+P_2+M_1+M_2$
is needed. But then the quantum reference system is $P_1+P_2+M_1+M_2$. 
One can see here explicitly 
the role of the fact that the internal state of $P_1+P_2+M_1+M_2$
usually cannot be reconstructed if the internal state of $P_1+M_1$ and that of
$P_2+M_2$ are given. 
When the measurements by $\tilde M_1$ and $\tilde M_2$ 
have been previously performed,
the internal state of $P_1+P_2+M_1+M_2$ is
\begin{eqnarray}
|\psi_{P_1+P_2+M_1+M_2}>=|\psi_{P_1+M_1}>|\psi_{P_2+M_2}>\;.\label{intst1}
\end{eqnarray}
In the absence of the extra measurements, however, it is
\begin{eqnarray}
|\psi_{P_1+P_2+M_1+M_2}>=\sum_j c_j|\phi_{P_1+M_1,j}>|\phi_{P_2+M_2,j}>
\neq |\psi_{P_1+M_1}>|\psi_{P_2+M_2}>.\label{intst2}
\end{eqnarray}
This difference is responsible for the violation of Bell's inequality,
while all the interactions take place locally. 

Summarizing, we have seen that the initial internal state
of $P_1$ ($P_2$) determines the first (second) measurement
process, therefore, these states 'carry' the initial correlations
and 'transfer' them to the measuring devices. 
As the measurement processes do not influence each other, 
the observed correlations may stem only from the 'common past'
of the particles.
On the other hand, any attempt to
compare the initial internal states of $P_1$ and $P_2$ with
the results of both measurements changes the correlations,
thus a joint probability for the simultaneous existence of these states
cannot be defined. This means that the reason for the
 violation of Bell's inequality is that the usual derivations
 always assume that the states (or 'stable properties')
 which carry the initial correlations can be freely compared with the results
 of the measurements. This comparability is usually 
 thought to be a consequence of realism.
 According to the present theory, the above assumption
 goes beyond the requirements of realism and proves to be wrong,
 because each of the states $|\psi_{P_1+M_1}>$, 
$|\psi_{P_2+M_2}>$, $|\psi_{M_1}>$ and $|\psi_{M_2}>$ exists individually,
but they cannot be compared without changing the correlations.

\section{Conclusion}

A new theoretical framework for nonrelativistic quantum mechanics has
been presented. It coincides with the usual one (especially, 
Schr\"odinger's
equation is not modified), except for the postulates
concerning the measurement. Instead of these latter, a dependence 
of the states
on quantum reference systems (they are themselves 
physical systems) is postulated. The relations among 
the different kinds of states have 
been consistently postulated allowing one to recover all the
usual predictions concerning the experiments and also to resolve 
the well known old paradoxes (Schr\"odinger's cat paradox and the EPR
paradox). It has also been shown that despite of the violation of 
Bell's inequality correlations between separated systems can always be
attributed to a previous interaction ('common past'), i.e., quantum
mechanics is a local theory. An important feature of the present approach 
is that 
measurements and macroscopic systems do not play a privileged role. 
Measurements are considered as usual interactions corresponding to 
some special
Hamiltonians. An {\em a priori} classical background is also absent.
 In principle this renders possible to check the validity 
 of the present approach, 
 as the observed macroscopic properties, first of all,
the localization of the internal states of macroscopic bodies 
in both coordinate and momentum 
space should follow from the theory when
using realistic assumptions concerning the relevant energies, 
structure and interactions. Demonstrating this would be a convincing
argument in favour of the present approach. (Note that it would also 
demonstrate that
classical mechanics can be derived from quantum mechanics.)
 The solution of this
question needs, however, a lot of further work and is beyond the 
scope of the
present paper.

Summarizing, the achievements of the theory are the following:\hfill\break
        {\bf i,} it gives a foundation of 
quantum mechanics independently of classical physics\hfill\break
        {\bf ii,} it is free from inconsistencies\hfill\break
        {\bf iii,} it gives in principle 
concrete and well defined predictions concerning 
        the possible results of a measurement 
        and their probabilities, 
provided the initial quantum state of the measuring device 
        and the measured object is given.\hfill\break
        {\bf iv,} it resolves the famous paradoxes (Schr\"odinger's cat and EPR), and\hfill\break
        {\bf v,} explains why the violation of Bell's inequality
        does not imply nonlocality.

Let us consider now the question how the
the dependence on the usual coordinate frames enters the present 
formalism,
or, more generally, how the canonical transformations appear.
To this end, let us remark first that we considered all the time
nonrelativistic quantum mechanics, and therefore all the physical 
systems has been assumed to consist of a given number of
particles which never decay. Let us perform some unitary 
transformations 
in the Hilbert spaces of all the particles. It is easy to see that
the whole formalism presented in Section 3 is covariant with 
respect to 
these transformations, i.e, the same relations hold 
among the transformed quantities as have previously held 
among the untransformed ones. (Note that the Hamiltonian will not be
invariant in general, unless in the special case of symmetry 
transformations.)
The reason is that the operations occuring in the {\em Postulates} 
(i.e., the trace
operation and the calculation of eigenvalues and eigenvectors of 
density
matrices) are themselves covariant. Hence the present approach has 
the 
same covariance properties with respect to unitary transformations
as traditional quantum mechanics. 

As for the link to traditional coordinate systems let us consider 
some group of geometric symmetry (e.g. the rotation group)
\footnote{As is
well known, such groups have unitary representations in Hilbert 
spaces.} 
and a multitude of isolated systems $I_j$
which will play the role of coordinate systems. Suppose that each 
$I_j$ has 
a distinguished subspace in its Hilbert space such a way that each 
state 
can be transformed by the application of a suitable group element 
into a state
lying in this subspace, while any state in the subspace leaves the 
subspace
under the application of any group element. 
The state of $Q$ with respect to the quantum reference system
         $Q+I_j$ coincides with the internal state of $Q$, as $I_j$
        is isolated. This state, however, may be still transformed
        by any element of the symmetry group to get an equivalent
        description. Let us choose that group element 
        which transfers the internal state of $I_j$ into its 
        distinguished subspace. The state of $Q$ we get this way
        is what one can call the description of the
        quantum system $Q$ with respect to the coordinate system $I_j$.

Finally, let us emphasize, that according to the present theory,
there is no collapse of the wave function, there are 
no parallel worlds, no nonlocal interactions, no corrections
     contributing to Schr\"odinger's equation if it is applied to
        macroscopic systems, and gravity has nothing to do with 
        quantum measurements. There is, however, a new principle, the 
        dependence
        of the states on quantum reference systems, 
        which removes inconsistencies from quantum mechanics 
        and which might serve
        as a guide if one wants to establish quantum theory in a new field.

\section{Acknowledgements}

The author is indebted to A.Bringer, G.Eilenberger, M.Eisele, 
R.Graham, G.Gy\"orgyi, F.Haake, H.Lustfeld, P.Rosenqvist,
 P.Sz\'epfalusy, Z.Kaufmann, and G.Vattay for useful discussions 
 and remarks,
 to P.Sz\'epfalusy also for a critical reading of the manuscript
and for valuable comments.
 The author wants to thank for the hospitality of the {\em Institut f\"ur 
 Festk\"orperphysik, Forschungszentrum J\"ulich GmbH} where a 
 substantial 
 part of the work has been done. The thorough and constructive
 work of the referees and the editor is particularly thanked and acknowledged.
 This work has been partially supported by the Hungarian Academy of 
 Sciences
 under Grant Nos. OTKA T 017493, OTKA F 17166 and OTKA F 019266.
 The present paper came into being 
 within the framework of a scientific and technological cooperation agreement
 between the Hungarian and the German government 
 as a result of a research cooperation supported
 by the OMFB (Hungary) and the BMFT (Germany).

\begin{appendix}
\section{Proof of {\em Proposition 1}}

Let us
expand $|\psi_{A+B}>$ (cf. Eq.(\ref{e10})) 
in terms of the possible internal states 
$|\phi_{B,j}>$. (It is possible, as these
latter constitute a complete orthonormed set of states. ) We get
\begin{eqnarray}
|\psi_{A+B}>=\sum_j d_j |\zeta_j>|\phi_{B,j}>\quad,
 \label{e12}
\end{eqnarray}
where $|\zeta_j>$-s are normed states in the Hilbert space of
$A$, and $d_j$-s are appropriate coefficients. We prove that the
$|\zeta_j>$-s coincide with the possible internal states
$|\phi_{A,j}>$. Indeed, calculating $\hat \rho_B(A+B)$ (by using 
Eq.(\ref{e12})
 and {\em Postulate 4}) one obtains 
\begin{eqnarray}
\hat \rho_B(A+B)\nonumber\\
=\sum_j \sum_k d_j^* d_k <\zeta_j|\zeta_k>
|\phi_{B,k}><\phi_{B,j}|,
 \label{e13}
\end{eqnarray}
which must coincide with $\sum_j
|\phi_{B,j}>\lambda_j<\phi_{B,j}|$ (here $\lambda_j$ stands for
the $j$-th eigenvalue of $\hat \rho_B(A+B)$), cf. {\em Definition 3}. 
Therefore, we get
\begin{eqnarray}
<\zeta_j|\zeta_k>=\delta_{j,k}\quad,
 \label{e14}
\end{eqnarray}
and $|d_j|^2=\lambda_j$. Calculating now $\hat \rho_A(A+B)$ we
get 
\begin{eqnarray}
\hat \rho_A(A+B)=\sum_j |\zeta_j>|d_j|^2<\zeta_j|\quad,
 \label{e15}
\end{eqnarray}
which, in view of Eq.(\ref{e14}) shows that 
$|\phi_{A,j}>=|\zeta_j>$.

As a final remark, suppose that $|\xi_j>$ and $|\chi_j>$ are arbitrary
orthonormed set of states in the Hilbert space of $A$ and of
$B$, respectively, then expanding Eq.(\ref{e10}) on the basis 
$|\xi_l>|\chi_m>$ ($l,m=1,2,3,...$) one gets
\begin{eqnarray}
\gamma_{l,m}=\sum_j c_j \alpha_{j,l} \beta_{j,m}\quad.
 \label{e11}
\end{eqnarray}
Here $\gamma_{l,m}=\left(<\xi_l|<\chi_m|\right)|\psi_{A+B}>$,
$\alpha_{j,l}=<\xi_l|\phi_{A,j}>$, and
$\beta_{j,m}=<\chi_m|\phi_{B,j}>$. The two latter matrices
are obviously unitary ones. If $l$ and $m$ is restricted
to be smaller than some $N>>1$, then Eq.(\ref{e11}) is the well known
{\em singular value decomposition}\cite{sing} of the general 
complex matrix
$\gamma_{l,m}$.

\section{Correspondence between the Copenhagen interpretation and
the present approach}

In a special case (cf. the discussion following Eq.(\ref{e47}) it has 
already been shown that the state one obtains when using the idea of 
the collapse of the wave function corresponds in the present approach 
to the situation when the reference system is chosen to be the 
complementer system of the measuring device. We derive here this 
result
in more general terms. Suppose that we are given a microscopic 
system $Q$
and a measuring device $M$. They are initially separated and 
the compound system $Q+M$ is isolated all the time. One can see that 
we are
still using an idealized model, although it is well known that 
quantum states of macroscopic systems
are extremely sensitive (due to their enormous level density), thus
they cannot be practically isolated. For a correct description one 
should
include in $M$ a large neighbourhood of the measuring device - this, 
however,
does not modify our conclusion. The reason is that the result 
of the measurement is a rather robust property, not sharing the 
sensitivity
of the corresponding quantum state. This is why our idealization 
is acceptable.

Suppose that the measurement on a subsystem $S$ of $Q$ has been
performed. Denoting by $|\xi_{S,j}>$ the eigenstates of the measured 
quantity, the change of the state of the whole system $Q+M$ can be
expressed as
\begin{eqnarray}
|m_0>\sum_j c_j |\xi_{S,j}>|\chi_{Q\setminus S,j}>\nonumber\\
\rightarrow
\sum_j c_j |m_j>|\xi_{S,j}>|\chi_{Q\setminus S,j}>,\label{apa1}
\end{eqnarray}
where $|m_j>$ refers to the possible internal states of the measuring
device (this follows from the orthogonality of the states
$|\xi_{S,j}>$, cf. {\em Proposition 1}), 
while $c_j |\chi_{Q\setminus S,j}>$-s are 
(not necessarily orthogonal) 'coefficient states' arising
when expanding the initial internal state of $Q$ in terms of
the states $|\xi_{S,j}>$. The Copenhagen interpretation tells us
that due to the collapse of the wave function only one of the terms 
of the sum in Eq.(\ref{apa1}) survives, implying that the states of 
$M$,
$Q$, $S$, and $Q\setminus S$ are $|m_j>$, 
$|\xi_{S,j}>|\chi_{Q\setminus S,j}>$, $|\xi_{S,j}>$, and 
$|\chi_{Q\setminus S,j}>$, respectively. According to the present
interpretation, the possible internal states of $Q$ are just the
$|\xi_{S,j}>|\chi_{Q\setminus S,j}>$-s (due to the orthogonality 
of the states $|m_j>$), thus, if the $j$-th possible result has been
observed, the states 
$|\xi_{S,j}>|\chi_{Q\setminus S,j}>$, $|\xi_{S,j}>$, and 
$|\chi_{Q\setminus S,j}>$ may be identified with $\hat \rho_Q(Q)$, 
$\hat \rho_S(Q)$ and $\hat \rho_{Q\setminus S}(Q)$, respectively. 
Note that
all the latter states are dyads. In conclusion, we have seen that
the states obtained using the concept of the collapse of
the wave function correspond in the present approach to a
specific choice of the reference system, namely, when
the reference system is the whole 
quantum system $Q$, whose subsystem is subject to the measurement.
\end{appendix}

\end{document}